\documentclass[manuscript]{acmart}
\AtBeginDocument{%
  }

\setcopyright{acmlicensed}
\copyrightyear{2018}
\acmYear{2018}
\acmDOI{XXXXXXX.XXXXXXX}
\acmConference[Conference acronym 'XX]{Make sure to enter the correct
  conference title from your rights confirmation email}{June 03--05,
  2018}{Woodstock, NY}
\acmISBN{978-1-4503-XXXX-X/2018/06}

\usepackage[most]{tcolorbox}

\newtcbox{\rhl}{enhanced,
  boxrule=0pt,
  colback=yellow!40,
  arc=1mm,
  auto outer arc,
  boxsep=1pt,
  left=2pt,
  right=2pt,
  top=1pt,
  bottom=1pt,
  on line,       
  nobeforeafter  
}
\usepackage{graphicx}
\usepackage[table,xcdraw]{xcolor}
\usepackage{float}
\usepackage{enumitem}
\usepackage{tabularx}
\usepackage{booktabs}
\usepackage{csquotes}
\usepackage{xltabular}
\usepackage{multirow}
\usepackage{fancyhdr}
\usepackage{textcomp}
\usepackage{makecell}
\usepackage[a-1b]{pdfx}
\usepackage{array}
\newcolumntype{P}[1]{>{\raggedright\arraybackslash}p{#1}}
\usepackage[export]{adjustbox}

\begin{document}

\title[Technology Caregiving]{Technology Caregiving: Reframing How Older Adults Are Supported in Everyday Digital Activities}

\author{Debaleena Chattopadhyay}
\affiliation{%
  \institution{University of Illinois Chicago}
  \streetaddress{851 S. Morgan}
  \city{Chicago}
  \state{Illinois}
  \country{USA}
  \postcode{60607}
}
\email{debchatt@uic.edu}

\author{Tasneem Mubashshira}
\affiliation{%
  \institution{University of Illinois Chicago}
  \streetaddress{851 S. Morgan}
  \city{Chicago}
  \state{Illinois}
  \country{USA}
  \postcode{60607}
}
\email{tmubas2@uic.edu}


\begin{abstract}
Transformed by digitization, everyday activities---paying bills, shopping, managing transportation---increasingly require older adults to navigate digital systems. To accomplish these digital activities of daily living (DADLs), older adults often rely on help that looks less like IT support---institutional, episodic, and product-oriented---and more like caregiving: relational, ongoing, and aimed at preserving their functional independence. We argue that this practice is \textit{technology caregiving} and introduce a framework characterizing it along four dimensions: \textit{why} support is needed, \textit{who} provides it, \textit{when} it occurs, and \textit{how} it is delivered. Applying this framework, we then systematically review the literature on how older adults are supported in DADLs. From 3,381 unique records, 36 articles met the inclusion criteria. Findings show that technology caregiving involves burden, like traditional care, but is distinctly shaped as much by digital systems and their constant change as by technology caregivers' and older adults' abilities.

\end{abstract}



\maketitle

\section{Introduction}

As global populations age, the ability to age in place---maintaining independence within one’s own home and community---has become a central focus of aging research and public policy \cite{UN_WorldSocialReport_2023, WHO_AgeingHealth_2024}. Traditionally, the benchmarks for such independence have been the successful performance of activities of daily living (ADLs), such as feeding, dressing, and mobility \cite{Katz_1963_ADL}; instrumental activities of daily living (IADLs), such as managing medications, preparing meals, and handling personal finances \cite{Lawton_Brody_1969_IADL}; and enhanced activities of daily living (EADLs), such as pursuing hobbies or lifelong learning \cite{Rogers_2020_EADL}. As digital technologies become more deeply embedded in the fabric of everyday life, many of these activities are now carried out through digital systems \cite{Pew_2024_Mobile_Broadband, mendoza2024digitalization}, and carrying them out requires a further set of tasks that exist only because of that shift: creating accounts, configuring devices, connecting to networks, and keeping credentials current \cite{mois2024developing}. We adapt the term digital activities of daily living (DADLs) \cite{mois2024developing} to refer to both—the everyday activities now performed digitally and the tasks that make performing them possible. 



Just as difficulty with ADLs can lead to institutionalization \cite{luppa2010prediction}, difficulty with DADLs now risks a distinct form of functional exclusion. For example, the inability to operate a banking app or log into a patient portal is becoming as much a barrier to healthy aging as the inability to manage finances or medication. Older adults may find DADLs challenging \cite{sharifi2023b_seniorlearning, sharifi2023a_techcaregiving}, whether because of age-related physical or cognitive changes \cite{Kebede_2022_DigitalEngagementReview, Schroeder_2023_MappingReview}, a lack of digital confidence \cite{an2022understanding, bertolazzi2024barriers}, or feature-rich interfaces that change constantly \cite{yu2024reducing, yu2023history, lazar2026interrogating}.

Much like those who need assistance with ADLs or IADLs, older adults who struggle with DADLs often rely on help—from people, such as friends, family, and community helpers \cite{sharifi2023b_seniorlearning, sharifi2023a_techcaregiving, sharifi_development_2024, pang2021technology, olsson_warm_2018, geerts2024exploring, geerts2023bridging}; from automated tools, such as search engines and chatbots \cite{Pradhan_2020_VoiceAssistants_OlderAdults, li2018understanding}; or from a combination of the two. This kind of help is commonly called information technology (IT) support (or tech support). In its standard sense, IT support is the product-oriented provision of assistance---typically institutional and paid---that resolves specific technical problems with a device, application, or system \cite{rockart1983management, brancheau1993management}. It is organized around discrete incidents and delivered by personnel whose role is defined by the product or the institution behind it, rather than by any relationship with the user. For instance, when an instructor cannot configure a learning management system (LMS), the university's IT service desk or the LMS vendor's tech support team fixes the problem and closes the ticket.

We argue that helping older adults navigate DADLs, though often labeled tech support, is a fundamentally different practice. It is organized not around isolated incidents but around a \textit{person}. It is ongoing, relational, and aimed at preserving independence rather than restoring a product to working order. Recent work in gerontology has begun to recognize this, naming those who assist with DADLs, digital caregivers or digital support partners and observing that they may be informal, such as friends and family, or formal, such as professionals with technical expertise \cite{mois2024developing}. We turn to the practice itself and make the case that such help is better understood as \textit{care} work than as IT support---a practice we call \textit{technology caregiving}. Like traditional care, it is frequently invisible, unpaid labor carried out by family and friends \cite{sharifi2023b_seniorlearning, sharifi2023a_techcaregiving}. Distinguishing technology caregiving from IT support is not merely terminological: it reframes what effective help looks like, and therefore what interventions, resources, and systems we should design to support it.

Not all older adults need help with technology, and needing it is not unique to them. Older adults sometimes help others with everyday digital activities \cite{geerts2024exploring, sharifi2023a_techcaregiving, sharifi2023b_seniorlearning, hunsaker2020unsung}. Much research has studied technology design for supporting older adults \cite{xing2026tangibletopics,sixsmith2013technologies,munteanu2019designing} and their traditional caregivers \cite{hsu2024dancing,theopilus2025digital}. A recent review of HCI research on care found that care was studied in three main areas---managing health conditions, supporting everyday living, and sustaining caregivers---and in each, technology was the instrument of care, but using it was not an activity that itself called for care \cite{wang2026caring}. Supporting older adults in their everyday digital activities has only recently drawn attention in HCI \cite{pang2021technology,sharifi_development_2024, sharifi2023a_techcaregiving, sharifi2023b_seniorlearning}, where it remains scattered across the field under many labels, not yet examined as a coherent whole. While prior systematic reviews have addressed particular tool use among older adults, such as conversational AI or mobile health apps, and specific support types, such as digital literacy programs \cite{huang2025designing, Lee_2023_mHealth_CaregiverCommunication, Schroeder_2023_MappingReview, xu2025systematic, Gates_WilsonMenzfeld_2022_GeragogyReview,theopilus2025digital}, none has synthesized the assistance that DADLs demand: how support, whether human or computational, is distributed across the different stages of technology use, who provides it, and how it reaches them.

Our contributions are threefold. First, we distinguish the practice of helping older adults with everyday digital activities from IT support, and argue that it is better understood as care work—technology caregiving. Although the HCI literature has so far described this practice as a variant of IT support, we show that it is often ongoing, relational, and person-oriented, rather than episodic, tiered, and product-oriented. Second, we introduce the technology caregiving framework, which characterizes the practice along four dimensions: \textit{why} support is needed (motivational, instructional, remedial, or delegation), \textit{who} provides it (formal, informal, or self-reliant sources), \textit{when} it occurs (across stages of technology use, from pre-adoption to continued use), and \textit{how} it is delivered (structured or unstructured; collocated, remote, or hybrid). Across these dimensions, support is \textit{mediated} by artifacts, such as physical tools, computational tools, and social relationships. Third, we apply this framework in a systematic qualitative review of 36 papers drawn from seven databases and supplementary citation searching. Synthesizing this literature, we characterize the support older adults received for DADLs as configurations of purpose, source, timing, and delivery, together with the artifacts through which that support reached them. We further identify two themes across those configurations: the subjective and objective burden of technology caregiving, and the role of the digital environment in shaping what support was needed, who could provide it, when, and how.

\section{Background}

To situate our work, we review the literature at the intersection of aging, digital technology use, and caregiving. We first define digital activities of daily living (DADLs), then describe some of the specific challenges older adults encounter with digital systems while acknowledging the diversity in their levels of technology adoption, use, and confidence in use, and finally,  we discuss related work within the broader caregiving literature.

\subsection{Digital Activities of Daily Living (DADLs)}

In gerontology, functional independence is measured by an individual's ability to perform activities of daily living (ADLs), fundamental self-care tasks such as bathing, toileting, and transferring in and out of bed \cite{Katz_1963_ADL}. Instrumental activities of daily living (IADLs) cover the more complex tasks of independent community living, such as shopping, keeping house, and arranging transportation \cite{Lawton_Brody_1969_IADL}. Enhanced activities of daily living (EADLs) extend the category to additional activities that sustain social engagement, such as planning travel or volunteering in community organizations \cite{Rogers_2020_EADL}.

IADLs and EADLs are increasingly performed on digital platforms rather than through physical and analog interactions \cite{mois2024developing, sharifi2023b_seniorlearning}. Tasks once completed on paper forms or in person now require navigating mobile applications and web interfaces \cite{wagner2012review, prakash2013older, liu2016smart, mendoza2024digitalization}. Managing patient portals, coordinating rideshare services, and banking online are performed through digital systems rather than merely assisted by them \cite{wu2021older, maguire2024study, kruse2020utilization, almeida2017older}. This shift has also generated a second set of tasks with no analog predecessor---creating and recovering online accounts, configuring devices, connecting to networks, and keeping credentials and authentication methods current \cite{mois2024developing}---tasks that are means to an end, yet without which the activities that depend on them cannot proceed. We adapt the term digital activities of daily living (DADLs) \cite{mois2024developing} to refer to both. Digital skills such as operating a mouse, executing a touchscreen gesture, or typing do not themselves constitute a DADL, just as grip strength and balance do not constitute an ADL; nor does digital literacy, such as recognizing online fraud or identifying misinformation, just as health and financial literacy do not constitute an IADL. But creating, recovering, and maintaining online account credentials or \textit{using} online search to evaluate the credibility of a news article is a DADL. (Henceforth, we use DADLs and everyday digital activities interchangeably.)

Whereas ADLs concern physical capability, DADLs concern the ability to navigate a digital environment that is often unfamiliar and frequently updated \cite{sharifi2023b_seniorlearning, pang2021technology, mahmud2020learning}. When an older adult cannot file taxes online or manage multi-factor authentication, they may lose access to services they need, and that loss limits their independence in the same way that being unable to complete ADLs or IADLs does \cite{mois2024developing, Benge_2023_DigitalMethodsIADL, zhu2025unlocking, knowles2021harm, money2024barriers}. Aging in place therefore depends not only on personal digital confidence \cite{an2022understanding} but also on the availability of quality support \cite{sharifi2023b_seniorlearning, sharifi2023a_techcaregiving}.

\subsection{Older Adults and Digital Technology Use}

The definition of an ``older adult'' is not fixed to a chronological age \cite{hutchison2010life, vines2015age}; published thresholds range from 50 and above to 70 and older \cite{wilson2021using, piper2016understanding, tong2016serious}. Aging is often \textit{situational}: a person may not identify as \textit{old} in general but may feel so in relation to particular environments, objects, or rapid change \cite{brandt2010communities}. We adopt an anti-deficit model of aging, which avoids treating older users as a \textit{problematic other} or aging as inherently disabling \cite{knowles2021harm}, and instead recognizes older adults as active, discriminating users who evaluate technology against perceived risk, personal values, and cultural expectations \cite{bhattacharjee2020older, sharifi2023a_techcaregiving, knowles2018wisdom}.

Unlike younger cohorts, however, older adults had no formative exposure to the smartphones, tablets, chatbots, and online services that now define daily life; these became widely available only later in their lives \cite{hutchison2010life,ivan2020impact}. Technology use in later adulthood is also shaped by long-standing social roles, gender expectations, and cultural norms \cite{Schroeder_2023_MappingReview,ivan2020impact,antonio2014gender}: some older adults adapt new tools to fit familiar routines, while others resist change to maintain continuity with those roles \cite{atchley1989continuity}. This lack of formative exposure contributes to a nuanced landscape of technology adoption. Device ownership is high---most older adults now own a smartphone---but use is often limited in breadth and depth \cite{AARP_2025_TechTrends50Plus,piper2016understanding,Schroeder_2023_MappingReview}: they use fewer applications, and fewer features within them, than younger users \cite{piper2016understanding, li2018understanding, li2016older, sharifi2023b_seniorlearning}. They are selective, ignoring tools they judge nonessential or misaligned with their values \cite{knowles2018wisdom,vines2015age,waycott2016not}, and notably resistant to simplified, senior-friendly variants \cite{waycott2016not, knowles2021harm}. Because many do not view themselves as disabled, they reject tools designed for ``the elderly,'' which can perpetuate negative stereotypes and discourage the collaborative, intergenerational experiences that mainstream technology affords \cite{brandt2010communities, knowles2021harm, sharifi2023a_techcaregiving}.

Even when the intention to use digital applications for essential tasks such as health management or transportation is strong, older adults face substantial usability barriers \cite{sharifi2023b_seniorlearning}. Feature-rich interfaces hide functions behind layered navigation and complex menus, and older users tend to focus on central screen content rather than peripheral cues such as tab menus or side drawers \cite{li2020older, li2018understanding}. Many are also diffident about self-exploration, less comfortable than younger users with trial-and-error \cite{an2022understanding, mahmud2020learning}; when they cannot locate a feature, they may become stuck, repeating the same incorrect steps and giving up on the task \cite{yu2023history, yu2024reducing, yu2020maps}. These barriers are compounded by constant change in digital systems, where frequent updates to visual layouts and interface components can disrupt established mental models \cite{yu2024reducing, gilbert1999age, lazar2026interrogating}. Because navigating this shifting environment demands continual relearning and sustained confidence, older adults' transition to DADLs often remains incomplete or stressful \cite{sharifi2023b_seniorlearning, sharifi2023a_techcaregiving}. A comprehensive review of usability barriers and age-friendly design is, however. beyond the scope of this paper.

\subsection{Caregiving and Aging}

As older adults face physical or cognitive barriers in managing everyday activities, caregivers have long emerged as essential figures in sustaining their functional independence and well-being \cite{Wolff_2016_JAMAIM_Caregivers,Freedman_2014_MilbankQ_DisabilityCare}. Their role usually centers on ADLs, such as mobility and hygiene \cite{Riffin_2017_JAGS_UnpaidCaregivers,Muramatsu_2019_MedicaidHomeCareAides}, and IADLs, such as managing finances and medication \cite{Freedman_2014_MilbankQ_DisabilityCare,Havens_2024_ProxyAccounts,Wolff_2016_JAMAIM_Caregivers}. Caregivers are typically distinguished by compensation, training, and relationship to the care recipient: informal or family caregivers are generally unpaid but bound by personal relationships, offering emotional ties and familiarity \cite{Riffin_2017_JAGS_UnpaidCaregivers,Spillman_2014_MQ_NSOC}, whereas formal or professional caregivers, such as physicians, nurses, or home health aides, are paid or volunteer workers with specialized medical or caregiving skills \cite{Muramatsu_2019_MedicaidHomeCareAides, stone2019future}. Caregivers shape not only older adults' health and safety but also their ability to engage meaningfully with their community \cite{Marasinghe_2022_AssistiveDevices_BMCGeriatr,Moreno_2024_Gerontechnologies_AgingInPlace}. Intergenerational mentoring, in which younger individuals assist older adults informally or professionally, is often cited as a particularly effective form of such support \cite{Gates_WilsonMenzfeld_2022_GeragogyReview,xu2025systematic}. Because caregiving is both critical to an older adult's quality of life and often demanding for those who provide it, researchers have long sought to reduce this labor through technology-mediated interventions \cite{zhou2024technology,madara2016assistive}. 

These interventions include remote monitoring systems \cite{caldeira2023compare,duranvega2019iot}, communication systems \cite{chen2016effect,hwang2020exploring,neves2015hand}, and safety alarms \cite{fernandez2024technological,Choukou_2023_DHT_Caregivers_Scoping}; educational applications to reduce caregiver stress and improve care delivery \cite{FullerTyszkiewicz_2020_JMIRMH_CaregiverAppRCT,zhou2024technology}; and, more recently, artificial intelligence (AI) systems for care recipient monitoring and decision-making \cite{Borna_2024_AI_CaregiverSupport,milella2023artificial}. A comprehensive review of such tools for the physical or domestic care of older adults, however, is beyond the scope of this paper. Instead, we argue in the following section that the digitization of daily life has created a new, distinct form of labor: helping older adults with everyday digital activities.

\section{Technology Caregiving}

In this section, we argue that helping older adults navigate everyday digital activities is better understood as care work than as IT support \cite{rockart1983management, brancheau1993management}. We begin with Fisher and Tronto's conceptualization of care as an ongoing process \cite{fisher1990toward, tronto1998ethic}, and use its vocabulary to define technology caregiving as a distinct practice and compare it to both IT support and traditional caregiving. Finally, we situate technology caregiving within HCI research.

\subsection{A Definition of Caregiving}

Semantically, care is both an activity and a disposition. Of these, we adopt the perspective from the ethics literature, where care is defined as an activity that includes everything aimed at ``maintaining, continuing, or repairing the world'' \cite{fisher1990toward}. Caring involves taking the concerns and needs of the other as the basis for action---but caring work can be done without a caring disposition. While care involves some form of ongoing connection, it also involves conflict, such as care recipients having different ideas about their needs than their caregivers do. Care consists of four analytically distinct but interconnected phases \cite{fisher1990toward}. Caring about is noticing a need and deciding it warrants a response. Taking care of involves assuming responsibility for an identified need and determining how to respond to it. Care-giving involves the direct, hands-on work of meeting a need, and typically brings the caregiver into contact with the object of care. Care-receiving is the recipient's response to the care provided, the only reliable indication of whether the need was actually met. Finally, care involves accepting some form of burden that often depends on the availability of resources, including material goods, time, and skills \cite{fisher1990toward, pearlin1990caregiving, braithwaite1992caregiving}.

In the gerontology literature, caregiving has been conceptualized as a specific type of social support—although there is no broader consensus on the point at which support becomes caregiving \cite{chappell2011social}. Some define it as support ``provided to seniors because their health has deteriorated and they can no longer function independently in areas where they previously did'' \cite{chappell1992social}. Others impose no such decline requirement, defining care as ``the physical, mental and emotional activities and effort involved in looking after, responding to, and supporting others,'' \cite{baines1998women} similar to the activity-based definition of care we adopted above \cite{fisher1990toward}. However, across most definitions, caregiving is considered sustained rather than episodic labor, provided by someone trustworthy and reliable—whether a professional, paid worker (a formal caregiver) or a family member, friend, neighbor, or community member (an informal caregiver) \cite{chappell2011social, noelker1989home}. A third type, less commonly distinguished, is self-care—the care older adults provide for themselves \cite{penning2000self, penning1990self, genet2011home}.

\subsection{Defining Technology Caregiving}
Helping older adults complete their everyday digital activities has been studied across disciplines, from applied health sciences \cite{bertolazzi2024barriers} and gerontology \cite{mois2024developing, pizzul2025implementing} to, most recently, HCI \cite{pang2021technology, sharifi2023b_seniorlearning}. Across these fields, this help is sometimes described as a resource, such as a help desk \cite{lee2021factors}, chatbot \cite{yu2024reducing}, or set of printed instructions \cite{du2021lessons}, and sometimes an activity, such as training \cite{ahmad2022effectiveness}, troubleshooting \cite{bhattacharjee2020older}, or completing a task as a proxy \cite{foong2025prevalence}. Terms such as information and communications technology (ICT) support \cite{geerts2024exploring, tsai2019senior}, IT support \cite{leung2012older}, technology support \cite{sharifi2026helping}, warm support \cite{olsson_warm_2018}, and tech help \cite{jiang2026help} are used interchangeably to name this help. Yet neither framing accounts for how thought and action are interrelated toward maintaining older adults' quality of life and preserving their functional independence, nor for the ongoing, relational nature of the help involved. We argue that this help is better understood as a \textit{practice}. 

More specifically, it is a practice of \textit{care.} Helping older adults with everyday digital activities shares all the characteristics of care work described above. The basis for this action is almost always the older adult's need for functional independence and quality of life \cite{sharifi2023a_techcaregiving}, rather than the repair of a device or the maintenance of an IT service. Even when help comes from professionals or institutions, library and senior-center staff care about the older adult's continued participation in daily life \cite{geerts2023bridging, geerts2024exploring}, and instructors in community settings are frequently older adults themselves, helping peers out of shared investment, not institutional mandate \cite{pihlainen2021perceived, pizzul2025implementing}. In this, they parallel traditional formal caregivers such as nurses, home health aides, and volunteer community advocates \cite{chappell2008comparing, lan2002subcontracting}.

Older adults, in turn, seek help from those they find reliable and trustworthy---a familiar store, a community member, or a grandchild \cite{sharifi2023a_techcaregiving, sharifi_development_2024}. Like traditional caregiving, this help is relational and ongoing rather than episodic, and so it can strain relationships: it draws on the helper's resources of time, skill, and patience, and it can generate conflicting views between older adults and their helpers about what help is needed, how it should be provided, and whether it meets the need \cite{sharifi2023a_techcaregiving}. Finally, older adults also help themselves---using paper notebooks to remember passwords, or digital tools such as AI chatbots and online search to locate a setting \cite{sharifi_development_2024}. By the source of the help, then, we can identify three types—formal, informal, and self-reliant support---mirroring the same types of traditional caregivers \cite{penning2000self, penning1990self, genet2011home}. Helping older adults with everyday digital activities is therefore not incidental to care but an instance of it. We call it \textit{technology caregiving}, and define it as follows:

\begin{quote}
    Technology caregiving is the practice of supporting older adults in navigating the digital activities of daily living—the ongoing, relational work that maintains their quality of life and preserves their functional independence.
\end{quote}

Yet technology caregiving is a distinctive instance of care. What sets it apart from traditional caregiving is the environment in which it takes place: digital systems change frequently and iterate rapidly, so the work is rarely finished \cite{lazar2026interrogating}. Its success depends not only on the abilities of the caregiver and the older adult but on systemic conditions neither controls---internet and device access, and platform choices made by third parties, such as whether a provider adopts a Cerner or an Epic patient portal \cite{sharifi2023a_techcaregiving}. Needs therefore recur not only because the person's circumstances change but because the technology itself does.

If the digital environment marks one boundary of technology caregiving, IT support marks the other. By our definition, IT support is a type of technology caregiving---the formal type---when it is oriented to the older adult's independence rather than to a device \cite{geerts2024exploring, geerts2023bridging}. A help-desk agent who resets a locked account and closes the ticket is repairing a service; one who, in the same call, checks whether the older adult can now sign in on their own and shows them how to avoid the lockout next time is helping them stay independent. Even when so oriented, the activity may be carried out without a caring disposition, meeting the need mechanically. Table~\ref{tab:caregiving-comparison} compares the three along these lines.

\newcolumntype{L}{>{\raggedright\arraybackslash}X} 
\newcolumntype{R}{>{\raggedleft\arraybackslash}X}  
\newcolumntype{K}[1]{>{\raggedleft\arraybackslash}p{#1}}  

\begin{table}[ht]
\centering
\caption{A comparison of technology caregiving, traditional caregiving, and IT support.}
\label{tab:caregiving-comparison}
\begin{tabularx}{\textwidth}{
    R   
    P{3.2cm}          
    P{3.7cm}           
    P{2.5cm}           
}
\toprule
\textbf{} & \textbf{Technology Caregiving} & \textbf{Traditional Caregiving} & \textbf{IT Support} \\
\midrule
\textbf{Scope} &&&\\
Basis for action & Older adult's needs & Older adult's needs &  Product or service\textsuperscript{1} \\
Activities supported & DADLs & ADLs / IADLs (non-digital) & DADLs\textsuperscript{1} \\

\textbf{Care markers} &&&\\
Relational & Yes & Yes & No\textsuperscript{1} \\
Ongoing (vs.\ episodic) & Yes & Yes & No\textsuperscript{1} \\
Basis of trust & Relational & Relational & Transactional\textsuperscript{1} \\
Involves burden & Yes & Yes & No\textsuperscript{1} \\
Measure of success & Needs met & Needs met & Task completed\textsuperscript{1} \\

\textbf{Organization} &&&\\
Sources of help & Formal, informal, self & Formal, informal, self & Formal only \\
Tiered & No & No & Yes \\
Paid & Sometimes & Sometimes & Always \\
Institutional & Sometimes & Sometimes & Almost always \\

\textbf{Digital environment} &&&\\
Digital technology dependence & Yes & No & Yes \\
Rate of infrastructure change & Fast & Slow & Fast \\
\bottomrule
\addlinespace[2pt]
\multicolumn{4}{@{}p{\textwidth}@{}}{\textsuperscript{1}\, \footnotesize{IT support qualifies as a formal type of technology caregiving when it is oriented to the older adult's needs rather than a product or service; on the marked rows, it then matches the technology-caregiving column.}} \\
\end{tabularx}
\end{table}

\subsection{Situating Technology Caregiving in HCI}

Although we bring the vocabulary of care to the study of IT support, support for technology adoption and use is not itself new to HCI. It has long been part of adoption models for older adults---a lineage running from the Unified Theory of Acceptance and Use of Technology (UTAUT) at the turn of the millennium \cite{venkatesh2003user}, through its adaptation for older adults in the Senior Technology Acceptance and Adoption Model (STAM) \cite{renaud2008predicting}, to later ease-of-learning and exploration models \cite{barnard2013learning, tsai2019senior, sharifi2023b_seniorlearning}. Across this lineage, support consistently appears as a facilitating condition that shapes whether an older adult experiments with, and ultimately adopts, a technology. Yet these models conceptualize support as a resource rather than a practice, and largely as a monolith: instruction manuals, computer training, help desks, and encouragement from family are all folded into a single construct, drawing no distinction between institutional and social support, nor between the different ways each is enacted. This framing reflects the setting in which the models were developed: technology use was then studied primarily within institutions—workplaces, universities, and government organizations—where treating support as institutional was consistent with how the technology itself was encountered \cite{rockart1983management, brancheau1993management}.

The digitization of everyday life is, by contrast, a recent development, and for older adults it decouples technology use from any single institution \cite{mois2024developing}. Much of their technology use now unfolds outside the workplace, in the ordinary settings of daily life, and it is this shift that surfaces a more care-oriented account of support---one the resource-and-institution framing cannot fully accommodate. This does not render traditional IT support obsolete; rather, the appropriate form of support depends on the digital activity in question. Specialized or workplace activities---a clinician documenting a visit in an electronic health record, say---remain well served by institutional IT support, whereas everyday digital activities such as banking, managing health online, and video calling are better served by technology caregiving. For older adults' everyday digital activities, then, what these adoption models call support is technology caregiving.

Situating technology caregiving in relation to social support draws a second connection. Like traditional caregiving, technology caregiving is a specific type of social support \cite{chappell1992social}, and the boundary between them is often blurred, just as it is in the gerontology literature \cite{chappell2011social}. In the social support literature, the perceived availability of support is associated with older adults' well-being \cite{taylor2011social}; technology caregiving inherits this relational character, in which perceived support can bolster an older adult's confidence and willingness to engage with technology \cite{yu2023history}. What sets technology caregiving apart from social support is the digital environment that is central to the practice---an environment often changing without warning, which continually reshapes the activities older adults must perform and alters the resources on which caregivers must draw. Next, we characterize the practice: how it comes together across the stages of technology use, who provides it, and how it addresses the needs that arise.

\section{A Framework of Technology Caregiving}

\subsection{Methods}

Having argued that helping older adults with everyday digital activities is a practice of care, we now set out to characterize that practice. We do so by developing a conceptual framework following a \textit{theory-adaptation} approach to theory development, in which an existing conceptualization is reframed through another theoretical lens to propose a novel perspective on it \cite{jaakkola2020designing}. In theory-adaptation terms, IT support is the \textit{domain theory}—the existing body of knowledge about the phenomenon—and care is the \textit{method theory}, the lens through which it is reframed, one we showed to be more fitting than technology adoption and social support (Section 3.3). This reframing yielded technology caregiving as a distinct conceptual domain, which we now characterize along four dimensions.

We derived these dimensions from the essential constituents of care work (distinct from the phases of care discussed in Section 3.1). Any act of care takes another's needs as the basis for action, is provided by someone, unfolds over time, and is enacted in some manner; it therefore has a purpose, a source, a timing, and a mode of delivery, which we frame as \textit{why} support is needed, \textit{who} provides it, \textit{when} it occurs, and \textit{how} it is delivered. For each dimension, we followed a typology approach, specifying its categories by integrating the prior concepts and theories that best fit the facet \cite{jaakkola2020designing}. Categories were reinterpreted and revised where recent empirical findings did not fit. Two authors independently mapped the recurring categories for each dimension and reconciled their mappings through discussion, refining the categories into the mutually distinct types this approach calls for.

\subsection{Framework}

\subsubsection{The Purpose}

The why dimension captures the purpose of support: what the older adult needs to accomplish everyday digital activities. We adapt an empirically derived taxonomy of IT support needs among older adults \cite{geerts2024exploring}, taking up its account of needs but not of the sources that meet them. This yields four types of purpose for technology caregiving: \textit{motivational}, \textit{instructional}, \textit{remedial}, and \textit{delegation}. Motivational support fosters awareness, confidence, and the willingness to engage at all, e.g., when a daughter encourages her father to try video calling. Instructional support shows how to carry out an activity that is new to the older adult, e.g., following a video tutorial to learn to send a text message. Remedial support addresses an activity that has broken, changed, or otherwise gone wrong, e.g., when a community-center volunteer repairs smartphone settings after a system update. Delegation involves another party carrying out the digital activity, or part of one, on the older adult's behalf, e.g., when a grandson creates a social media account that the older adult then uses on their own.


\begin{figure}[t]
\centering
 \includegraphics[width=.55\linewidth]{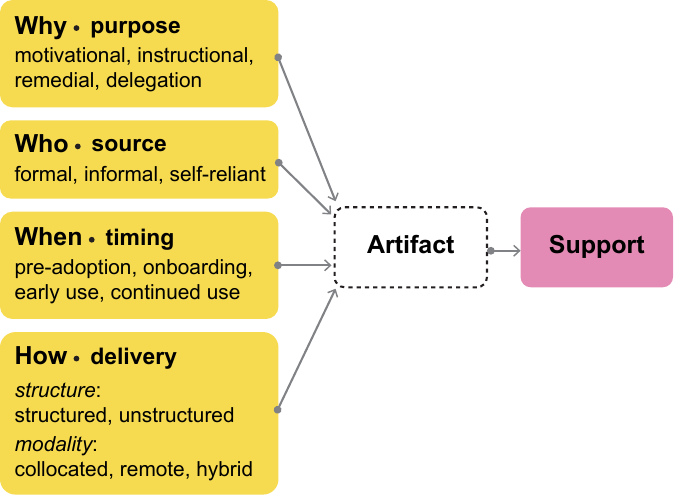} 
    \caption{The Technology Caregiving Framework: Technology caregiving is characterized along four dimensions—\textit{why} support is needed, \textit{who} provides it, \textit{when} it occurs, and \textit{how} it is delivered, in terms of both structure and modality. Any technology caregiving episode has all four at once, and the support reaches the older adult through a mediating artifact.}
    \label{fig:framework}
\end{figure}

We depart from this taxonomy in two ways \cite{geerts2024exploring}. First, we set aside its split between instructional support and technical support, which that taxonomy defined as system-repair work. Instruction and repair are difficult to disentangle within a single activity \cite{sharifi2023b_seniorlearning, jiang2026help, pang2021technology}, and when a familiar activity is disrupted, the help needed is the same in kind whether the \textit{system} breaks or the \textit{skill} does \cite{sharifi2023b_seniorlearning, sharifi2023a_techcaregiving}. We therefore define remedial support by what the help addresses---a breakdown to a familiar activity---rather than by whether the fix is system- or skill-related. This also narrows instructional support to activities that are new to the older adult, rather than \textit{any} learning of a digital skill. Second, delegation is broader than the taxonomy's proxy use: we define it as the transfer of responsibility for an activity, or part of one, so that complete proxy use is one end of a range rather than the category itself \cite{sharifi2023a_techcaregiving, sharifi2023b_seniorlearning}---a son may do all of his mother's online banking while she uses social media on her own. Throughout, these categories name the primary need that \textit{prompted} the technology caregiving episode: remedial support may incidentally teach a new skill, and a delegated activity may be partly instructional. Appendix~\ref{app:framework} discusses both departures in more detail.

\subsubsection{The Source}
The who dimension identifies the source of support. We define its types directly from those already established for traditional caregiving of older adults: \textit{formal}, \textit{informal}, and \textit{self-reliant}. Formal support comes from a professional, paid or unpaid, without a prior personal relationship to the older adult---an IT specialist, a librarian, or an instructor at a senior center. Informal support comes from someone with whom the older adult has an existing personal relationship---a family member, friend, neighbor, or fellow community member---and is typically unpaid. Self-reliant support is what an older adult does for themselves, without another person's involvement in the moment, relying instead on an artifact: a paper notebook with steps written for checking a bank balance, an online video tutorial, a search engine, or an AI chatbot \cite{sharifi_development_2024, sharifi2023b_seniorlearning}. It is an edge case for technology caregiving, since caregiving is by definition a relational practice usually provided by another party, but we include it because older adults sometimes meet their own needs in ways that mirror the self-care already recognized in the traditional caregiving literature \cite{penning2000self, penning1990self, genet2011home}. Formal and informal support are mediated by artifacts too, as we discuss in Section 4.2.5.

\subsubsection{The Timing}

The when dimension captures the stage of technology use during which technology caregiving occurs. We build its categories on the models of technology adoption and use discussed in Section 3.3 \cite{venkatesh2003user, barnard2013learning,renaud2008predicting, sharifi2023b_seniorlearning}, yielding four stages: \textit{pre-adoption}, \textit{onboarding}, \textit{early use}, and \textit{continued use}. Pre-adoption is the stage in which intention is formed, shaped by the older adult's perceptions and attitudes and by their social environment. Onboarding covers the initial setup that makes a technology usable, e.g., when a store employee transfers applications and settings from an old smartphone to a new one. Early use is the period of experimentation that follows, when the older adult begins working with a limited set of functions, e.g., calling and texting on that new smartphone, before incorporating the technology into daily life. Continued use is the stage at which the technology has been accepted and has become part of the older adult's routine.

We depart from earlier models \cite{venkatesh2003user, renaud2008predicting, barnard2013learning} in dividing their account of experimentation and exploration into two stages, because setting up a technology and beginning to use it involve different work \cite{pang2021technology, sharifi2023b_seniorlearning}. Onboarding is bounded and often one-time---migrating accounts, transferring settings, installing and configuring a device---and calls on skills an older adult may never need again, whereas early use is the repeated, exploratory work of discovering what a technology can do. The support each calls for differs accordingly.

Transitions between these four stages are neither linear nor one-directional: they can be progressive or regressive, and are often dictated by the presence or absence of effective technology caregiving \cite{sharifi2023a_techcaregiving, sharifi2023b_seniorlearning, sharifi_development_2024}. An older adult who abandons a technology returns to pre-adoption rather than leaving the framework, since taking it up again requires that intention be formed anew before onboarding or early use can follow---prompted, perhaps, by a need the technology now answers, such as checking test results in a patient portal while waiting for an appointment.

\subsubsection{The Delivery}

The how dimension captures two aspects: how support is organized (structure) and how it reaches the older adult (modality). Support can be \textit{structured} or \textit{unstructured}. Structured support is planned and deliberate, often arranged in advance for the purpose of providing help---a technology class at a senior center or a workshop on obtaining health insurance online. Unstructured support is casual and spontaneous, arising in the course of ordinary interaction---a granddaughter answering a question over dinner. What varies along this aspect is how deliberately the help is organized, not who provides it: a phone store employee answering a passing question is unstructured support from a formal source, while a relative recording a step-by-step tutorial video is structured support from an informal one.

We build the categories for structure on the distinction between formal, nonformal, and informal learning in the lifelong learning literature \cite{la1982formal}, setting aside its boundary between formal and nonformal instruction, which depends on whether the instruction leads to a credential, which few older adults pursue for everyday digital activities, and has little bearing on technology caregiving. We therefore call organized support of either kind structured, and incidental support unstructured. These labels also avoid confusion with the who dimension, where formal and informal already denote the source of support.

The modality of support delivery can be \textit{collocated}, \textit{remote}, or \textit{hybrid}. Collocated support is given with both parties physically present. Remote support is given at a distance, and may be synchronous, as in a phone or video call, or asynchronous, as in an exchange of messages. Hybrid support combines the two within a single technology caregiving episode, as when a video call is followed by written steps sent via email. This distinction parallels the one between collocated and remote healthcare, where telehealth has reshaped when and how care reaches older adults \cite{csahin2024telemedicine}. Support in any modality may be structured or unstructured, a scheduled video demonstration is structured, an unprompted text message is not.

\subsubsection{The Mediating Artifact}

The four dimensions describe what support is for, who provides it, when it occurs, and how it is delivered. None of them, however, names what the support is carried by. We call this the \textit{artifact} that mediates technology caregiving: whatever the support works through. An artifact in this sense need not be a physical object, or an object at all. Nor is it a fifth dimension. Each dimension sorts an episode into one of a closed set of mutually distinct categories, whereas the artifact names the means by which an episode is carried out. A single episode may run through several artifacts at once, and the types of artifact are an open list rather than an exhaustive one. We borrow the term from cultural-historical activity theory, where an action reaches what it is directed at through a mediating artifact rather than directly \cite{cole1993cultural, sannino2018}. Here, what the support is directed at is the older adult's accomplishment of the digital activity, and the artifact is what carries it there.

Artifacts can take many forms. Support may be mediated by something physical, such as a printed handout or a notebook of written steps; by something computational, such as a video tutorial or a screen-sharing application \cite{sharifi2023b_seniorlearning}; by the relationship through which help travels, since what a granddaughter offers is inseparable from her being there to ask \cite{sharifi2023a_techcaregiving}; by a professional's expertise, as when a librarian's familiarity with common account problems carries the help \cite{geerts2023bridging}; or by an established routine, such as a standing Sunday visit at which accumulated questions are worked through. What matters for the framework is that support is always carried by some means.

\subsection{Using the Framework}
Any episode of technology caregiving has a purpose, a source, a timing, and a manner of delivery at once, carried by an artifact (Figure~\ref{fig:framework}). Modality is the one exception: it presupposes two parties, so self-reliant support has none. A daughter helping her mother set up a rideshare account over dinner is providing informal, unstructured, collocated delegation at onboarding, mediated by the relationship that makes her easy to ask and by the sticky note on which she writes the login details---two artifacts in one episode. A home health aide's presentation about the benefits of a patient portal is formal, structured, collocated, motivational support at pre-adoption, mediated by handouts and by the aide's position of trust. An older adult who consults an AI chatbot after an online payment fails is meeting a remedial need during continued use through self-reliant, unstructured support, mediated by the chatbot alone.

The dimensions also shape which artifacts are used: formal sources often mediate through expertise and prepared materials, informal sources typically through relationships and improvised aids, and self-reliant support through physical and computational artifacts alone.

These configurations are analytic rather than fixed. A single technology caregiving episode may shift as it unfolds, as when instruction that fails gives way to delegation, or a scheduled session drifts into unplanned questions. What the framework offers is a vocabulary for describing these configurations and for asking which recur, which are effective, and which leave needs unmet.

Next, we use the framework as the analytic lens for a systematic review of the empirical literature on how older adults are supported in everyday digital activities. Of the questions the framework makes askable, we take up the first: which configurations recur in the evidence reported. Applying an a priori framework to a body of evidence, and refining it where the evidence does not fit, is an established way to evaluate a conceptual framework \cite{carroll2013best, carroll2011worked}. Because we derived the dimensions from literature distinct from the corpus, and fixed the framework before analyzing the corpus, the review tests what the conceptual argument alone cannot: whether the dimensions account for configurations we did not have in view, and whether together they describe the forms technology caregiving takes in practice.

\section{A Systematic Review of Technology Caregiving}

\subsection{Methods}

\begin{figure}[t]
\centering
 \includegraphics[width=.65\textwidth]{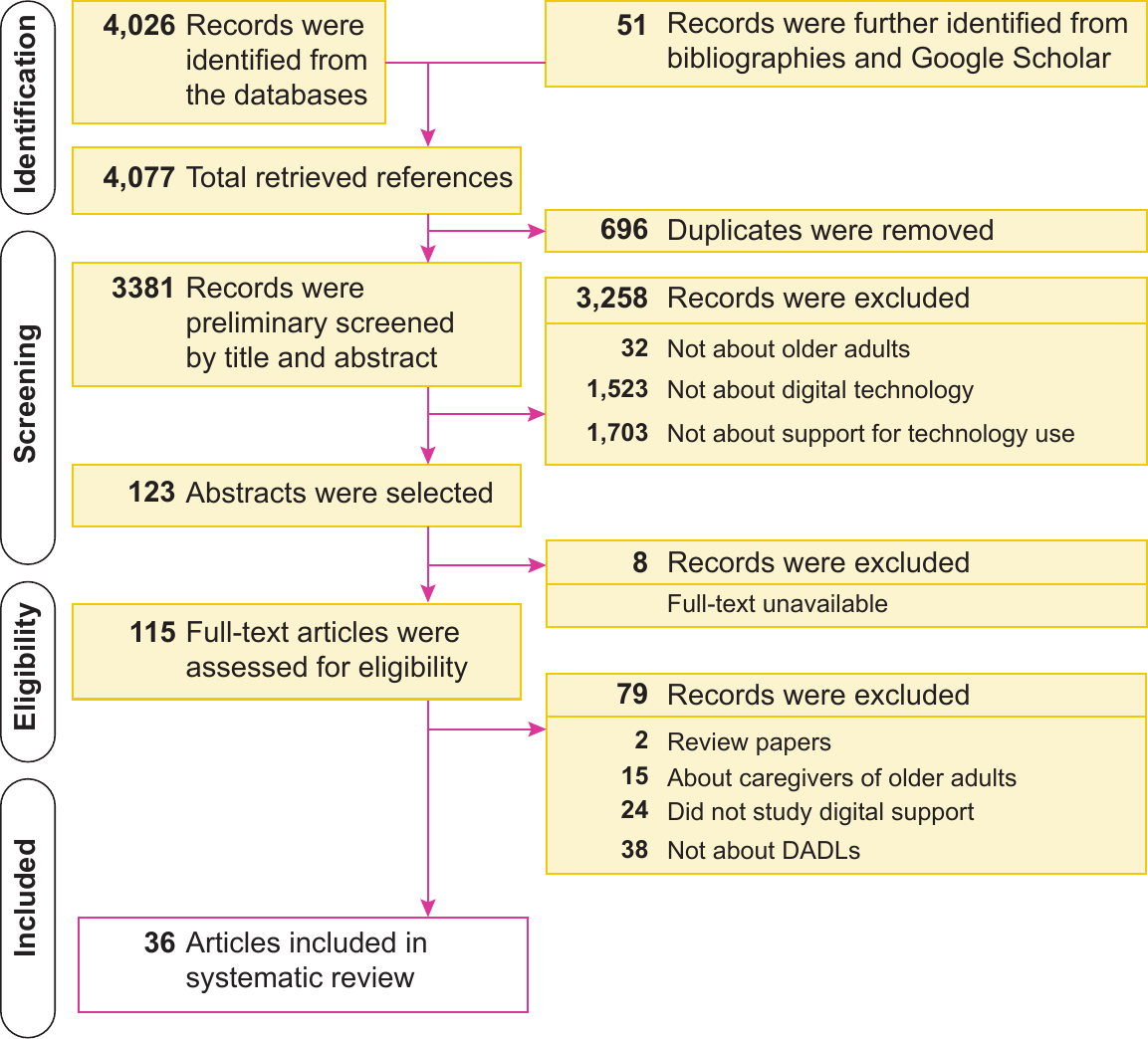} 
    \caption{Summary of literature search following PRISMA guidelines \cite{page2021prisma}.}
    \label{fig:prisma}
    \Description{A PRISMA flow diagram summarizing the literature search in four stages, running top to bottom, with excluded records branching to the right at each stage. Identification: 4,026 records were identified from the databases and 51 further records were identified from bibliographies and Google Scholar, giving 4,077 total retrieved references. Screening: 696 duplicates were removed, leaving 3,381 records that were preliminary screened by title and abstract; 3,258 records were excluded, comprising 32 not about older adults, 1,523 not about digital technology, and 1,703 not about support for technology use; 123 abstracts were selected. Eligibility: 8 records were excluded because the full text was unavailable, leaving 115 full-text articles that were assessed for eligibility; 79 records were excluded, comprising 2 review papers, 15 about caregivers of older adults, 24 that did not study digital support, and 38 not about DADLs. Included: 36 articles were included in the systematic review.}
\end{figure}

This systematic review followed the Preferred Reporting Items for Systematic Reviews and Meta-Analyses (PRISMA) guidelines \cite{page2021prisma}. We searched Google Scholar and seven databases spanning computing, health, and the behavioral sciences: Scopus, EMBASE, PsycINFO, IEEE Xplore, Cochrane Central Register of Controlled Trials, PubMed, and ACM Digital Library, targeting peer-reviewed, full-length, English-language publications from January 2015 through May 2026. Search queries combined database-specific controlled vocabulary with keywords spanning three primary domains---older adults, support, and digital technology---and were not restricted to any specific medical condition; the complete set of keywords is provided in the Appendix. We also manually searched the bibliographies of identified review articles using a snowballing approach \cite{sayers2008tips}. The literature search and selection process is summarized in the PRISMA diagram (Figure~\ref{fig:prisma}).

Studies were included if they reported an empirical investigation of support for older adults in accomplishing at least one digital activity of daily living (DADL; see Section 2.1), such as online shopping, social media use, or video calling. We excluded studies that (1) did not investigate how support directly helped older adults accomplish DADLs, (2) examined a non-digital activity assisted by a digital system, such as assembling a product with digital instructions, where removing the digital system would leave the activity intact, and only the support would disappear, (3) focused solely on physical or passive interventions (e.g., physical therapy, fall detection, or monitoring systems) that did not require the older adult's active digital engagement, or (4) examined digital tools used exclusively by caregivers without involving the older adult in the digital activity itself.

All retrieved records were managed in a Zotero library for deduplication and initial screening. The second author screened titles and abstracts against the inclusion and exclusion criteria to identify candidates for full-text assessment, referring uncertain records to the first author. Both authors then reviewed the full texts of the remaining candidates and discussed borderline cases, with the first author making the final eligibility decisions at both stages. For each eligible article, data were extracted into a structured spreadsheet: descriptive metadata (e.g., title, publication year), study details (e.g., sample size, target population), the types of DADLs examined (e.g., online banking, patient portal use), and details about the support described.

We analyzed the final corpus using thematic analysis in two forms \cite{braun2021one}. This synthesis had three aims. The first was evaluative: to test whether the framework's four dimensions account for the support configurations described in the literature, and to refine them where the evidence did not fit. The second was descriptive: to establish which configurations recur in the evidence reported. The third was comparative: to identify where technology caregiving resembles traditional caregiving and where it departs from it. We pursued the first two aims through codebook thematic analysis, with the technology caregiving framework (Figure~\ref{fig:framework}) serving as the a priori coding structure. The first author began by coding a subset of the corpus against the framework, extracting relevant findings and author interpretations while remaining open to material that fell outside its categories. The coding structure was then refined through group data analysis sessions, in which the framework's dimensions were tested against the evidence and the codebook was adjusted where studies did not fit, after which the first author coded the full corpus against the revised codebook. We pursued the third aim through reflexive thematic analysis, in which themes were developed interpretively rather than against a predetermined structure. Drawing on the coded extracts together with material that had fallen outside the framework's categories, the first author developed themes across the corpus. The research team met weekly to resolve ambiguities and  challenge interpretive assumptions.

\subsection{Results}

Our search retrieved 4,026 records from the databases, with 51 additional records identified from bibliographies and Google Scholar. After removing 696 duplicates, we screened the titles and abstracts of 3,381 records, of which 123 appeared to concern support for older adults' digital technology use; eight could not be retrieved, and the remaining 115 underwent full-text review, yielding 36 articles that met the inclusion criteria (Figure~\ref{fig:prisma}). At full-text review, articles were most often excluded because the supported activity was not a DADL (38), because they described older adults' technology use without examining support for it (24), because they concerned support for caregivers rather than for older adults (15), or because they were reviews (2).

We report the findings in three parts. We first describe the general characteristics of the included studies, including participant demographics and the DADLs examined (Table~\ref{tab:study-description}). We then map each study onto the technology caregiving framework (Table~\ref{tab:support}). Finally, we develop two broad themes that cut across the corpus and compare technology caregiving with traditional caregiving: feelings of burden among both technology caregivers and older adults; and the role of the digital environment.

\subsubsection{Study Characteristics}\textbf{Publication activity} in our corpus concentrated in recent years: three quarters of the 36 articles (75\%, $n = 27$) appeared between 2022 and 2026, with the final year only partially covered by our search. Most articles (72\%, $n = 26$) appeared in HCI and computing venues, with ACM CHI alone accounting for 13; the rest appeared in gerontology and health or in social science venues. These 36 articles reported 43 studies (Table~\ref{tab:study-description}). Data collection sites clustered in a small number of regions. Fourteen articles (39\%) reported data collected in East and Southeast Asia \cite{1_seo2023power, 2_jin2022synapse, 5_tang2022never, 9_jin2024exploring, 10_li2025explorar, 14_wan2019appmod, 17_fang2026guideme, 20_zhang2025wepilot, 22_wu2026doesn, 25_jin2022used, 27_hu2026not, 31_ma2023motivation, 33_gao2024easyask, 36_deng2025auntie} and twelve in North America \cite{3_mendel2019my, 6_tanprasert2024helpcall, 11_dai2025envisioning, 12_portz2019call, 16_hu2026looking, 18_hu2025surfacing, 19_piper2016technological, 21_wu2026providing, 28_haque2025envisioning, 29_garnett2026perceived, 30_finkelstein2023older, 32_zubatiy2021empowering}, together accounting for nearly three quarters of the corpus (72\%, $n = 26$); eight reported data collected in Europe \cite{7_hunsaker2019he, 8_luijkx2015grandma, 13_barber2025beyond, 23_zamir2018video, 24_soubutts2025hidden, 26_korpela2024investigating, 34_doyle2026factors, 35_petrovvcivc2024categorical}, and single articles in Australia \cite{4_kelly2024more} and Israel \cite{15_mendel2021exploratory}. No article in the corpus reported data collected in Africa, South America, or South Asia. Additional details about the study characteristics are reported in Appendix~\ref{app:corpus}.

\textbf{Participants.} The included studies drew on three kinds of data. Most (23 of 43 studies) collected data from older adults alone, fourteen from both older adults and technology caregivers, and the remaining six only from caregivers' accounts (Table~\ref{tab:study-description}); several of these described the older adults being helped in some detail, but that description came from the caregiver rather than from the older adult. Sample sizes ranged from five older adults in a co-design workshop \cite{13_barber2025beyond} to 701 respondents in a national survey \cite{35_petrovvcivc2024categorical}, but the studies in the corpus were predominantly qualitative with small samples: only four studies included more than 35 older adults \cite{7_hunsaker2019he, 8_luijkx2015grandma, 34_doyle2026factors, 35_petrovvcivc2024categorical}. Technology caregivers were usually younger than those they helped and were most often adult children and grandchildren \cite{4_kelly2024more, 5_tang2022never, 11_dai2025envisioning, 20_zhang2025wepilot, 36_deng2025auntie}; formal caregivers appeared less frequently and in institutionally specific roles \cite{21_wu2026providing, 22_wu2026doesn, 25_jin2022used, 29_garnett2026perceived, 34_doyle2026factors}. Several further characteristics were reported too unevenly to support corpus-wide claims, but the studies reporting them suggest a narrowing of who was studied: educational attainment skewed toward the highly educated, studies were split between recruiting older adults with cognitive impairment and screening them out, and prior technology experience was an eligibility requirement in several studies.

\textbf{Study design and instruments.} We categorized the corpus into two broad study types: \textit{generative} studies, which sought to understand needs and practices and reported empirical data about technology caregiving ($n = 31$), and \textit{evaluative} studies, which assessed a system, program, or intervention ($n = 12$; Table~\ref{tab:study-description}). Interviews were the most commonly used method, appearing in 22 of the 36 articles, often paired with think-aloud protocols or with co-design and workshop sessions.

\newcolumntype{L}{>{\raggedright\arraybackslash}X} 
\newcolumntype{R}{>{\raggedleft\arraybackslash}X}  
\newcolumntype{P}[1]{>{\raggedright\arraybackslash}p{#1}} 
\newcolumntype{K}[1]{>{\raggedleft\arraybackslash}p{#1}}  

\renewcommand{\arraystretch}{1}

\begin{xltabular}{\textwidth}{
    P{1cm}  
    c         
    l         
    P{2 cm}  
    P{2.5cm}  
    L         
}
\caption{Characteristics of the 36 reviewed articles, ordered by year of publication. Articles reporting more than one study are listed on separate rows, with sample sizes and age ranges corresponding to each study. Dashes indicate a participant group that was not included, or an age range that was not reported.}
\label{tab:study-description} \\
\toprule
\textbf{Article} & \textbf{Year} & \textbf{Study type} & \textbf{Older adults} (\textit{n}; age) & \textbf{Technology caregivers} (\textit{n}; age) & \textbf{DADL} \\
\midrule
\endfirsthead
\multicolumn{6}{l}{\textit{Table \thetable\ continued from previous page}} \\
\toprule
\textbf{Article} & \textbf{Year} & \textbf{Study type} & \textbf{Older adults} (\textit{n}; age) & \textbf{Technology caregivers} (\textit{n}; age) & \textbf{DADL} \\
\midrule
\endhead
\midrule
\endfoot
\bottomrule
\endlastfoot

\cite{8_luijkx2015grandma} & 2015 & Generative & 53; 65+ & -- & Several DADLs \\
\cite{19_piper2016technological} & 2016 & Generative & -- & 20; 27--56 & Several DADLs \\
\cite{23_zamir2018video} & 2018 & Evaluative & 18; 65+ & 9; -- & Making video calls \\
\cite{7_hunsaker2019he} & 2019 & Generative & 57; 59--91 & -- & Several DADLs \\
\cite{3_mendel2019my} & 2019 & Generative & -- & 86; 18+ & Managing online security and privacy \\
\cite{12_portz2019call} & 2019 & Generative & 24; 65+ & -- & Managing health online \\
\cite{14_wan2019appmod} & 2019 & Evaluative & 34; 50+ & 34; 18--40 & Managing online security and privacy \\
\cite{15_mendel2021exploratory} & 2021 & Generative & 18; 63--83 & -- & Managing online security and privacy \\
\cite{32_zubatiy2021empowering} & 2021 & Evaluative & 10; $M = 78$ & 10; $M = 73$ & Several DADLs \\
\cite{2_jin2022synapse} & 2022 & Evaluative & 18; 60--74 & -- & Several DADLs \\
\cite{25_jin2022used} & 2022 & Generative & 16; 60--79 & 7; -- & Managing online finances \\
\cite{5_tang2022never} & 2022 & Generative & 20; 61--88 & 18; 22--52 & Several DADLs \\
\cite{30_finkelstein2023older} & 2023 & Evaluative & 35; 55--90 & -- & Several DADLs \\
\cite{31_ma2023motivation} & 2023 & Generative & 30; 50--78 & 30; 50--78 & Several DADLs \\
\cite{1_seo2023power} & 2023 & Generative & 31; 60--69 & -- & Shopping online \\
\multirow{2}{=}{\cite{33_gao2024easyask}}
  & \multirow{2}{*}{2024} & Generative & 16; 56--76 & -- & \multirow{2}{=}{Several DADLs} \\
  &  & Evaluative & 18; 60--76 & -- &  \\
\multirow{2}{=}{\cite{9_jin2024exploring}}
  & \multirow{2}{*}{2024} & Generative & 18; 65--94 & -- & \multirow{2}{=}{Several DADLs} \\
  &  & Generative & 15; 65--94 & -- &  \\
\cite{4_kelly2024more} & 2024 & Generative & -- & 18; 20--76 & Making video calls \\
\cite{26_korpela2024investigating} & 2024 & Generative & 21; 65--89 & -- & Several DADLs \\
\cite{35_petrovvcivc2024categorical} & 2024 & Generative & 701; 65--85 & -- & Several DADLs \\
\cite{6_tanprasert2024helpcall} & 2024 & Evaluative & 14; 65--90 & -- & Several DADLs \\
\multirow{2}{=}{\cite{13_barber2025beyond}}
  & \multirow{2}{*}{2025} & Generative & 24; 60+ & 35; -- & \multirow{2}{=}{Managing online finances} \\
  &  & Generative & 5; 60--79 & 4; 50--59 &  \\
\cite{11_dai2025envisioning} & 2025 & Generative & 17; 65--84 & 10; 25--74 & Managing online finances \\
\cite{36_deng2025auntie} & 2025 & Generative & -- & 124; $Mdn = 26$ & Several DADLs \\
\multirow{2}{=}{\cite{28_haque2025envisioning}}
  & \multirow{2}{*}{2025} & Generative & 8; 62--75 & -- & \multirow{2}{=}{Using social media} \\
  &  & Evaluative & 12; 60+ & -- &  \\
\cite{18_hu2025surfacing} & 2025 & Generative & 6; 60s--80s & -- & Making video calls \\
\cite{10_li2025explorar} & 2025 & Evaluative & 18; 60+ & -- & Several DADLs \\
\cite{24_soubutts2025hidden} & 2025 & Generative & 7; 73--77 & -- & Several DADLs \\
\multirow{2}{=}{\cite{20_zhang2025wepilot}}
  & \multirow{2}{*}{2025} & Generative & 11; -- & 10; -- & \multirow{2}{=}{Several DADLs} \\
  &  & Evaluative & 12; 66--80 & 12; 22--32 &  \\
\cite{34_doyle2026factors} & 2026 & Evaluative & 146; 65--95 & 4; -- & Managing health online \\
\multirow{2}{=}{\cite{17_fang2026guideme}}
  & \multirow{2}{*}{2026} & Generative & 16; 55--86 & -- & \multirow{2}{=}{Several DADLs} \\
  &  & Evaluative & 18; $M = 63.6$ & -- &  \\
\cite{29_garnett2026perceived} & 2026 & Generative & 7; 64--95 & 32; 40--69 & Making video calls \\
\cite{16_hu2026looking} & 2026 & Generative & 10; 50s--80s & -- & Making video calls \\
\multirow{2}{=}{\cite{27_hu2026not}}
  & \multirow{2}{*}{2026} & Generative & -- & 20; 18--38 & \multirow{2}{=}{Several DADLs} \\
  &  & Generative & 16; 60--81 & -- &  \\
\cite{22_wu2026doesn} & 2026 & Generative & 10; 65--79 & 7; -- & Several DADLs \\
\cite{21_wu2026providing} & 2026 & Generative & -- & 20; 23--56 & Several DADLs \\

\end{xltabular}

\textbf{Digital activities of daily living.} Studies varied in how specific they were about which everyday digital activity they addressed (Table~\ref{tab:study-description}). Most did not isolate one: 21 of the 36 articles examined several DADLs at once, treating support as something that spans a person's digital life rather than attaching to a single activity. Across these articles, the activities that recurred most often were communicating through messaging and social media platforms \cite{2_jin2022synapse, 5_tang2022never, 19_piper2016technological, 20_zhang2025wepilot, 27_hu2026not, 33_gao2024easyask, 36_deng2025auntie}, shopping online \cite{2_jin2022synapse, 5_tang2022never, 6_tanprasert2024helpcall, 17_fang2026guideme, 27_hu2026not, 36_deng2025auntie}, managing money through banking and payment applications \cite{5_tang2022never, 24_soubutts2025hidden, 36_deng2025auntie}, and accessing health services \cite{5_tang2022never, 30_finkelstein2023older}. One article extended into government services, examining how older adults sustained access to digital welfare and pension systems \cite{22_wu2026doesn}. Across the studies conducted in China, WeChat recurred as a single application through which many of these activities were performed \cite{5_tang2022never, 20_zhang2025wepilot, 27_hu2026not, 31_ma2023motivation, 33_gao2024easyask, 36_deng2025auntie}. The remaining 15 articles examined a single DADL. Among them, making video calls was the most studied \cite{4_kelly2024more, 16_hu2026looking, 18_hu2025surfacing, 23_zamir2018video, 29_garnett2026perceived}, followed by managing online finances \cite{11_dai2025envisioning, 13_barber2025beyond, 25_jin2022used} and managing online security and privacy \cite{3_mendel2019my, 14_wan2019appmod, 15_mendel2021exploratory}, and managing health online \cite{12_portz2019call, 34_doyle2026factors}; shopping online \cite{1_seo2023power} and using social media \cite{28_haque2025envisioning} were each examined once. The three articles on security and privacy differed in what the activity consisted of: delegating app permission decisions \cite{14_wan2019appmod}, managing permission settings and passwords \cite{15_mendel2021exploratory}, and seeking help with security and privacy problems as they arose \cite{3_mendel2019my}.

\subsubsection{Support Configurations}

A category was recorded for an article when any of its studies provided evidence for it; the seven articles reporting two studies \cite{9_jin2024exploring, 13_barber2025beyond, 17_fang2026guideme, 20_zhang2025wepilot, 27_hu2026not, 28_haque2025envisioning, 33_gao2024easyask}, are listed as separate rows in Table~\ref{tab:study-description}. Support was overwhelmingly informal (30 of 36 articles), unstructured (28), directed at a technology already in continued use (34), and mediated by the social relationship between the older adult and the technology caregiver (31; Table~\ref{tab:support}). The recurring arrangement in the corpus was therefore a family member, friend, or neighbor helping without prior arrangement, during regular contact, with a device the older adult already owned or an application already installed. Twenty-six articles reported evidence in all four of these categories \cite{1_seo2023power, 3_mendel2019my, 4_kelly2024more, 5_tang2022never, 7_hunsaker2019he, 8_luijkx2015grandma, 11_dai2025envisioning, 12_portz2019call, 13_barber2025beyond, 14_wan2019appmod, 15_mendel2021exploratory, 17_fang2026guideme, 19_piper2016technological, 20_zhang2025wepilot, 22_wu2026doesn, 24_soubutts2025hidden, 25_jin2022used, 26_korpela2024investigating, 27_hu2026not, 28_haque2025envisioning, 29_garnett2026perceived, 30_finkelstein2023older, 31_ma2023motivation, 32_zubatiy2021empowering, 33_gao2024easyask, 36_deng2025auntie}. What separated one article from another was usually not a departure from this arrangement but an addition to it: a paid helper alongside the family one, an earlier stage of use, an expertise or a routine carrying part of the work.

Most articles reported more than one purpose. Delegation was the most frequently reported purpose (27 articles), followed by instructional support (25) and remedial support (22), with motivational support the least common (13). Only nine articles reported a single purpose: instructional support \cite{2_jin2022synapse, 6_tanprasert2024helpcall, 9_jin2024exploring, 33_gao2024easyask}, remedial support \cite{16_hu2026looking, 18_hu2025surfacing}, or delegation \cite{28_haque2025envisioning, 35_petrovvcivc2024categorical, 36_deng2025auntie}; seventeen reported three or more purposes and seven reported all four \cite{1_seo2023power, 5_tang2022never, 8_luijkx2015grandma, 12_portz2019call, 15_mendel2021exploratory, 19_piper2016technological, 24_soubutts2025hidden}. Motivational support never appeared on its own: every article that reported it also reported at least one other purpose, most often instructional support or delegation. The same type of technology caregiver typically served several purposes \cite{1_seo2023power, 3_mendel2019my, 4_kelly2024more, 5_tang2022never, 8_luijkx2015grandma, 12_portz2019call, 19_piper2016technological, 21_wu2026providing, 25_jin2022used, 29_garnett2026perceived, 34_doyle2026factors}, moving between them within an episode \cite{5_tang2022never, 7_hunsaker2019he, 19_piper2016technological, 25_jin2022used}.

Sources were more often singular. Nineteen articles reported one kind of source, twelve reported two, and five reported all three \cite{5_tang2022never, 7_hunsaker2019he, 15_mendel2021exploratory, 17_fang2026guideme, 25_jin2022used}. The largest group reported \textit{only} informal support (13 articles), which came from spouses, adult children, grandchildren, and friends \cite{1_seo2023power, 3_mendel2019my, 8_luijkx2015grandma, 12_portz2019call, 13_barber2025beyond, 19_piper2016technological, 24_soubutts2025hidden, 26_korpela2024investigating, 27_hu2026not, 31_ma2023motivation, 32_zubatiy2021empowering, 35_petrovvcivc2024categorical, 36_deng2025auntie}. Formal support was reported in thirteen articles but was seldom the only kind, appearing alongside informal support in ten of them. Formal support as the sole source was studied only in institutional settings---assisted living facilities \cite{21_wu2026providing}, care homes and a community hospital \cite{23_zamir2018video}, and a clinical trial with triage nurses \cite{34_doyle2026factors}. In all three, technology support was an additional responsibility taken on by staff employed to provide traditional caregiving.

This pattern held more broadly: in eight articles the formal source was someone employed for another purpose---bank employees \cite{11_dai2025envisioning, 25_jin2022used}, public service providers \cite{22_wu2026doesn}, and traditional caregivers \cite{4_kelly2024more, 29_garnett2026perceived}. Four articles described services that existed exclusively to support technology use: community technology training programs \cite{5_tang2022never}, a help desk \cite{7_hunsaker2019he}, mobile phone service centers \cite{17_fang2026guideme}, and a device-distribution program with training and an ongoing helpline \cite{30_finkelstein2023older}. Self-reliance was reported in fifteen articles, usually as one option among other types, such as informal support (7 articles) \cite{2_jin2022synapse, 6_tanprasert2024helpcall, 10_li2025explorar, 14_wan2019appmod, 20_zhang2025wepilot, 28_haque2025envisioning, 33_gao2024easyask}. Only three articles described older adults working through difficulties without another person present at all \cite{9_jin2024exploring, 16_hu2026looking, 18_hu2025surfacing}.

All but two articles reported support at the continued-use stage, and eleven reported it only there \cite{11_dai2025envisioning, 13_barber2025beyond, 14_wan2019appmod, 16_hu2026looking, 18_hu2025surfacing, 22_wu2026doesn, 27_hu2026not, 28_haque2025envisioning, 32_zubatiy2021empowering, 34_doyle2026factors, 35_petrovvcivc2024categorical}. Fewer articles reported support during early use (18), onboarding (15), and pre-adoption (10). Five articles reported support across all four stages \cite{1_seo2023power, 5_tang2022never, 8_luijkx2015grandma, 15_mendel2021exploratory, 21_wu2026providing}, and these reported more purposes than other articles did. Articles often fell into a single structure category, either unstructured (22) or structured (6); only six reported both \cite{4_kelly2024more, 6_tanprasert2024helpcall, 7_hunsaker2019he, 22_wu2026doesn, 25_jin2022used, 30_finkelstein2023older}. Among the six that reported only structured support, the form it took varied widely: prepared step-by-step instructional materials delivered through a computational tool that older adults could use independently \cite{2_jin2022synapse, 10_li2025explorar}; traditional caregivers providing support for a specific DADL on a fixed schedule, such as video calls \cite{23_zamir2018video} and managing health online \cite{34_doyle2026factors}; and older adults themselves organizing support in advance, such as writing call notes and keeping them within reach \cite{16_hu2026looking, 18_hu2025surfacing}. Most articles reported collocated support (24), and fourteen of these also reported remote support \cite{1_seo2023power, 3_mendel2019my, 4_kelly2024more, 5_tang2022never, 7_hunsaker2019he, 8_luijkx2015grandma, 11_dai2025envisioning, 12_portz2019call, 13_barber2025beyond, 15_mendel2021exploratory, 17_fang2026guideme, 24_soubutts2025hidden, 27_hu2026not, 30_finkelstein2023older}. Ten articles described collocated support only and six remote only. A single episode delivered partly at a distance and partly in person was reported just once \cite{4_kelly2024more}.

Artifacts are the means through which support is carried, not physical objects in the everyday sense (Section 4.2.5). We identified eleven kinds: social relationships (31 articles); expertise in the technology (11), in aging and older adults' care (6), in the DADL's domain, such as banking or health (5), and in a language the older adult did not read or speak (2); computational tools (13), physical tools (8), and additional devices (6); established routines (8); cultural norms (5); and physical proximity (6). The social relationship was the most common, and in six articles it was the only artifact described \cite{1_seo2023power, 12_portz2019call, 13_barber2025beyond, 26_korpela2024investigating, 32_zubatiy2021empowering, 35_petrovvcivc2024categorical}. Physical tools \cite{3_mendel2019my, 4_kelly2024more, 5_tang2022never, 6_tanprasert2024helpcall, 15_mendel2021exploratory, 16_hu2026looking, 18_hu2025surfacing, 25_jin2022used} included steps written down in a notebook, and established routines \cite{4_kelly2024more, 5_tang2022never, 16_hu2026looking, 18_hu2025surfacing, 22_wu2026doesn, 23_zamir2018video, 24_soubutts2025hidden, 36_deng2025auntie} a standing weekly visit at which accumulated issues were discussed. Cultural norms governing whether help could be asked for or offered were reported only in articles conducted in China \cite{5_tang2022never, 20_zhang2025wepilot, 22_wu2026doesn, 31_ma2023motivation, 36_deng2025auntie}. Most articles described support mediated through more than one of the eleven artifacts: twelve reported three, ten reported two, and eight reported four or more, including one that described seven \cite{22_wu2026doesn}.

\newcolumntype{L}{>{\raggedright\arraybackslash}X} 
\newcolumntype{R}{>{\raggedleft\arraybackslash}X}  
\newcolumntype{P}[1]{>{\raggedright\arraybackslash}p{#1}} 
\newcolumntype{K}[1]{>{\raggedleft\arraybackslash}p{#1}}  

\renewcommand{\arraystretch}{.9}

\begin{xltabular}{\textwidth}{
    P{2.8cm}  
    L         
}
\caption{Reviewed articles across the categories of each framework dimension and mediating artifacts. Articles are placed only where the study reports evidence for a category; an absence indicates unreported, not absent.}
\label{tab:support}\\
\toprule
\textbf{Support} & \textbf{Articles} \\
\midrule
\endfirsthead
\multicolumn{2}{l}{\textit{Table \thetable\ continued from previous page}} \\
\toprule
\textbf{Support} & \textbf{Articles} \\
\midrule
\endhead
\midrule
\endfoot
\bottomrule
\endlastfoot
\textbf{Purpose} & \\
Motivational & \cite{1_seo2023power,4_kelly2024more,5_tang2022never,8_luijkx2015grandma,10_li2025explorar, 12_portz2019call,15_mendel2021exploratory,19_piper2016technological,21_wu2026providing, 24_soubutts2025hidden,25_jin2022used,26_korpela2024investigating,34_doyle2026factors} \\
Instructional & \cite{1_seo2023power, 2_jin2022synapse, 3_mendel2019my,5_tang2022never, 6_tanprasert2024helpcall,7_hunsaker2019he,8_luijkx2015grandma, 9_jin2024exploring,10_li2025explorar, 12_portz2019call, 13_barber2025beyond,15_mendel2021exploratory,17_fang2026guideme,19_piper2016technological,20_zhang2025wepilot,22_wu2026doesn,24_soubutts2025hidden,25_jin2022used,26_korpela2024investigating,27_hu2026not,29_garnett2026perceived,30_finkelstein2023older,31_ma2023motivation,32_zubatiy2021empowering,33_gao2024easyask} \\
Remedial & \cite{1_seo2023power, 3_mendel2019my,4_kelly2024more,5_tang2022never,7_hunsaker2019he, 8_luijkx2015grandma,11_dai2025envisioning, 12_portz2019call,14_wan2019appmod,15_mendel2021exploratory,16_hu2026looking,17_fang2026guideme,18_hu2025surfacing,19_piper2016technological,20_zhang2025wepilot,21_wu2026providing,22_wu2026doesn,23_zamir2018video,24_soubutts2025hidden,29_garnett2026perceived,30_finkelstein2023older,34_doyle2026factors} \\
Delegation & \cite{1_seo2023power, 3_mendel2019my, 4_kelly2024more,5_tang2022never,7_hunsaker2019he,8_luijkx2015grandma,11_dai2025envisioning, 12_portz2019call,13_barber2025beyond, 14_wan2019appmod,15_mendel2021exploratory,17_fang2026guideme,19_piper2016technological,21_wu2026providing,22_wu2026doesn,23_zamir2018video,24_soubutts2025hidden,25_jin2022used,26_korpela2024investigating,27_hu2026not,28_haque2025envisioning,29_garnett2026perceived, 30_finkelstein2023older,31_ma2023motivation,32_zubatiy2021empowering,35_petrovvcivc2024categorical,36_deng2025auntie} \\
\midrule
\textbf{Source} & \\
Formal & \cite{4_kelly2024more,5_tang2022never,7_hunsaker2019he,11_dai2025envisioning,15_mendel2021exploratory,17_fang2026guideme,21_wu2026providing,22_wu2026doesn,23_zamir2018video,25_jin2022used,29_garnett2026perceived,30_finkelstein2023older,34_doyle2026factors} \\
Informal & \cite{1_seo2023power, 2_jin2022synapse,3_mendel2019my,4_kelly2024more,5_tang2022never,6_tanprasert2024helpcall,7_hunsaker2019he,8_luijkx2015grandma,11_dai2025envisioning,12_portz2019call,13_barber2025beyond,14_wan2019appmod,15_mendel2021exploratory,17_fang2026guideme,19_piper2016technological,20_zhang2025wepilot,10_li2025explorar,22_wu2026doesn,24_soubutts2025hidden,25_jin2022used,26_korpela2024investigating,27_hu2026not,28_haque2025envisioning,29_garnett2026perceived,30_finkelstein2023older,31_ma2023motivation,32_zubatiy2021empowering, 33_gao2024easyask,35_petrovvcivc2024categorical, 36_deng2025auntie} \\
Self-reliant & \cite{2_jin2022synapse,5_tang2022never,6_tanprasert2024helpcall,7_hunsaker2019he,9_jin2024exploring,10_li2025explorar,14_wan2019appmod,15_mendel2021exploratory,16_hu2026looking,17_fang2026guideme,18_hu2025surfacing,20_zhang2025wepilot,25_jin2022used,28_haque2025envisioning, 33_gao2024easyask} \\
\midrule

\textbf{Timing} & \\
Pre-adoption & \cite{1_seo2023power,4_kelly2024more,5_tang2022never,8_luijkx2015grandma, 12_portz2019call,15_mendel2021exploratory,21_wu2026providing,24_soubutts2025hidden,25_jin2022used,29_garnett2026perceived} \\
Onboarding & \cite{1_seo2023power,3_mendel2019my,4_kelly2024more,5_tang2022never,7_hunsaker2019he,8_luijkx2015grandma,15_mendel2021exploratory,19_piper2016technological,20_zhang2025wepilot,21_wu2026providing,23_zamir2018video,26_korpela2024investigating,29_garnett2026perceived,30_finkelstein2023older,36_deng2025auntie} \\
Early use & \cite{1_seo2023power, 2_jin2022synapse,5_tang2022never,6_tanprasert2024helpcall,7_hunsaker2019he,8_luijkx2015grandma,9_jin2024exploring,10_li2025explorar,15_mendel2021exploratory,17_fang2026guideme,19_piper2016technological,20_zhang2025wepilot,21_wu2026providing,24_soubutts2025hidden,25_jin2022used,30_finkelstein2023older,31_ma2023motivation,33_gao2024easyask} \\
Continued use & \cite{1_seo2023power,3_mendel2019my,4_kelly2024more,5_tang2022never,7_hunsaker2019he,8_luijkx2015grandma,9_jin2024exploring, 10_li2025explorar,11_dai2025envisioning, 12_portz2019call,13_barber2025beyond,14_wan2019appmod,15_mendel2021exploratory,16_hu2026looking,17_fang2026guideme,18_hu2025surfacing,19_piper2016technological,20_zhang2025wepilot,21_wu2026providing,22_wu2026doesn,23_zamir2018video,24_soubutts2025hidden,25_jin2022used,26_korpela2024investigating,27_hu2026not, 28_haque2025envisioning,29_garnett2026perceived,30_finkelstein2023older,31_ma2023motivation,32_zubatiy2021empowering,33_gao2024easyask,34_doyle2026factors,35_petrovvcivc2024categorical,36_deng2025auntie} \\
\midrule
\textbf{Structure} & \\
Structured & \cite{2_jin2022synapse, 4_kelly2024more, 6_tanprasert2024helpcall,7_hunsaker2019he,10_li2025explorar,16_hu2026looking,18_hu2025surfacing,22_wu2026doesn,23_zamir2018video,25_jin2022used,30_finkelstein2023older,34_doyle2026factors} \\
Unstructured & \cite{1_seo2023power,3_mendel2019my,4_kelly2024more,5_tang2022never, 6_tanprasert2024helpcall,7_hunsaker2019he,8_luijkx2015grandma, 11_dai2025envisioning, 12_portz2019call,13_barber2025beyond,14_wan2019appmod,15_mendel2021exploratory,17_fang2026guideme,19_piper2016technological,20_zhang2025wepilot,21_wu2026providing,22_wu2026doesn,24_soubutts2025hidden,25_jin2022used,26_korpela2024investigating,27_hu2026not,28_haque2025envisioning,29_garnett2026perceived,30_finkelstein2023older,32_zubatiy2021empowering,33_gao2024easyask,36_deng2025auntie,31_ma2023motivation} \\
\midrule

\textbf{Modality} & \\
Collocated & \cite{1_seo2023power,3_mendel2019my,4_kelly2024more, 5_tang2022never,7_hunsaker2019he,8_luijkx2015grandma,11_dai2025envisioning,12_portz2019call,13_barber2025beyond,15_mendel2021exploratory,17_fang2026guideme,19_piper2016technological,21_wu2026providing,22_wu2026doesn,23_zamir2018video,24_soubutts2025hidden,25_jin2022used,26_korpela2024investigating,27_hu2026not,29_garnett2026perceived,30_finkelstein2023older,31_ma2023motivation,32_zubatiy2021empowering,33_gao2024easyask} \\
Remote & \cite{1_seo2023power, 2_jin2022synapse,3_mendel2019my,4_kelly2024more, 5_tang2022never,6_tanprasert2024helpcall,7_hunsaker2019he,8_luijkx2015grandma,11_dai2025envisioning,12_portz2019call,13_barber2025beyond,14_wan2019appmod,15_mendel2021exploratory,17_fang2026guideme,20_zhang2025wepilot,24_soubutts2025hidden,27_hu2026not, 28_haque2025envisioning,30_finkelstein2023older,34_doyle2026factors} \\
Hybrid & \cite{4_kelly2024more}\\
\midrule

\textbf{Artifact} & \\
Social relationships & \cite{1_seo2023power, 2_jin2022synapse,3_mendel2019my,4_kelly2024more,5_tang2022never, 6_tanprasert2024helpcall,7_hunsaker2019he,8_luijkx2015grandma,11_dai2025envisioning,12_portz2019call, 13_barber2025beyond,14_wan2019appmod,15_mendel2021exploratory,17_fang2026guideme,19_piper2016technological,20_zhang2025wepilot,21_wu2026providing,22_wu2026doesn,24_soubutts2025hidden,25_jin2022used,26_korpela2024investigating,27_hu2026not,28_haque2025envisioning,29_garnett2026perceived,30_finkelstein2023older,31_ma2023motivation,32_zubatiy2021empowering,33_gao2024easyask,34_doyle2026factors,35_petrovvcivc2024categorical,36_deng2025auntie} \\
IT expertise & \cite{4_kelly2024more,7_hunsaker2019he,11_dai2025envisioning,15_mendel2021exploratory,17_fang2026guideme,22_wu2026doesn,23_zamir2018video,25_jin2022used,14_wan2019appmod,30_finkelstein2023older,31_ma2023motivation}\\
Domain expertise & \cite{28_haque2025envisioning,11_dai2025envisioning,22_wu2026doesn,25_jin2022used,34_doyle2026factors}\\
Aging expertise & \cite{4_kelly2024more,21_wu2026providing,23_zamir2018video,19_piper2016technological,29_garnett2026perceived,30_finkelstein2023older}\\
Multilingual expertise & \cite{15_mendel2021exploratory,21_wu2026providing}\\
Computational tools & \cite{2_jin2022synapse, 6_tanprasert2024helpcall,7_hunsaker2019he,9_jin2024exploring,10_li2025explorar,14_wan2019appmod,15_mendel2021exploratory,16_hu2026looking,17_fang2026guideme,18_hu2025surfacing,20_zhang2025wepilot,28_haque2025envisioning,33_gao2024easyask} \\
Physical tools & \cite{3_mendel2019my,4_kelly2024more,5_tang2022never, 6_tanprasert2024helpcall,15_mendel2021exploratory,16_hu2026looking,18_hu2025surfacing,25_jin2022used} \\
Established routines & \cite{4_kelly2024more, 5_tang2022never,16_hu2026looking,18_hu2025surfacing,22_wu2026doesn,23_zamir2018video,24_soubutts2025hidden,36_deng2025auntie} \\
Cultural norms & \cite{5_tang2022never,20_zhang2025wepilot,22_wu2026doesn,31_ma2023motivation,36_deng2025auntie}\\
Additional devices & \cite{4_kelly2024more,9_jin2024exploring, 10_li2025explorar,22_wu2026doesn,8_luijkx2015grandma, 7_hunsaker2019he}\\
Physical proximity & \cite{11_dai2025envisioning, 15_mendel2021exploratory,22_wu2026doesn,27_hu2026not,30_finkelstein2023older,33_gao2024easyask}\\

\end{xltabular}

Across the framework dimensions, three configurations recurred. Because a configuration requires its categories to occur together, we identified them within a study and counted the articles in which they appeared. First and most common was informal, unstructured support during continued use, mediated via social relationships (SC1; 26); in fourteen of these articles support was delivered in both collocated and remote form, and SC1 included all seven articles that reported all four purposes \cite{1_seo2023power, 5_tang2022never, 8_luijkx2015grandma, 12_portz2019call, 15_mendel2021exploratory, 19_piper2016technological, 24_soubutts2025hidden}. A second configuration was formal and structured (SC2; 7), most often in institutional settings, such as a care home, a bank, a public service counter, and a clinical trial involving older adults with multimorbidity \cite{4_kelly2024more, 7_hunsaker2019he, 22_wu2026doesn, 23_zamir2018video, 25_jin2022used, 30_finkelstein2023older, 34_doyle2026factors}. The third was self-reliant support through a computational tool (SC3; 13) \cite{2_jin2022synapse, 6_tanprasert2024helpcall, 7_hunsaker2019he, 9_jin2024exploring, 10_li2025explorar, 14_wan2019appmod, 15_mendel2021exploratory, 16_hu2026looking, 17_fang2026guideme, 18_hu2025surfacing, 20_zhang2025wepilot, 28_haque2025envisioning, 33_gao2024easyask}, which mostly provided instructional and remedial support and included four of the five articles that reported no role for social relationships \cite{9_jin2024exploring, 10_li2025explorar, 16_hu2026looking, 18_hu2025surfacing}. These configurations were not exclusive: eleven articles reported more than one \cite{4_kelly2024more, 7_hunsaker2019he, 14_wan2019appmod, 15_mendel2021exploratory, 17_fang2026guideme, 20_zhang2025wepilot, 22_wu2026doesn, 25_jin2022used, 28_haque2025envisioning, 30_finkelstein2023older, 33_gao2024easyask}, and one reported all three \cite{7_hunsaker2019he}.

\subsubsection{Burden in Technology Caregiving.}

Technology caregiving was reported as labor (19 articles) that caregivers more often perceived as burdensome (11) \cite{1_seo2023power, 5_tang2022never, 13_barber2025beyond, 19_piper2016technological, 20_zhang2025wepilot, 21_wu2026providing, 22_wu2026doesn, 25_jin2022used, 29_garnett2026perceived, 32_zubatiy2021empowering, 36_deng2025auntie} than as rewarding (3) \cite{19_piper2016technological, 21_wu2026providing, 31_ma2023motivation}; seven articles reported it without characterizing it \cite{3_mendel2019my, 4_kelly2024more, 6_tanprasert2024helpcall, 7_hunsaker2019he, 11_dai2025envisioning, 24_soubutts2025hidden, 27_hu2026not}. Seeking or receiving support was likewise more often fraught with a negative feeling among older adults (22) \cite{1_seo2023power, 2_jin2022synapse, 5_tang2022never, 6_tanprasert2024helpcall, 7_hunsaker2019he, 8_luijkx2015grandma, 9_jin2024exploring, 11_dai2025envisioning, 13_barber2025beyond, 15_mendel2021exploratory, 17_fang2026guideme, 20_zhang2025wepilot, 22_wu2026doesn, 25_jin2022used, 26_korpela2024investigating, 27_hu2026not, 28_haque2025envisioning, 29_garnett2026perceived, 30_finkelstein2023older, 31_ma2023motivation, 33_gao2024easyask, 36_deng2025auntie}, such as shame, embarrassment, anxiety, fear of judgment, or the sense of being a burden, than a positive feeling (4) \cite{8_luijkx2015grandma, 12_portz2019call, 21_wu2026providing, 26_korpela2024investigating}, such as pride, eagerness, gratitude, or enrichment. We did not consider burden that lay in using the technology itself rather than in giving or receiving support for it. An absence of reluctance to ask was not construed as positive affect. Articles used a variety of terms for this labor, including burden, effort, workload, cost, and stress.

\textbf{Informal helpers.} Helpers described similar burdens across sources of support. What differed was whether their role set any limit on the work and whether anyone else could take it on. For informal helpers the support landed on top of everything else they were already doing. \textit{``I work two jobs and take care of my father, along with trying to have a social life, and I have two children of my own. So I'm overwhelmed''} (adult child, \cite{19_piper2016technological}). It came without warning, in the middle of other work: \textit{``Sometimes when I am in a meeting, my phone buzzes with a WeChat message from my mom, asking `How do I ...?' Moments like this can be distracting''} (adult child, \cite{20_zhang2025wepilot}). And it recurred without resolving---\textit{``I teach them once, and they don't get it; then I teach them again, and still, they don't get it''} (younger family member, \cite{20_zhang2025wepilot}). What then prompted a helper to stop teaching and do the task instead was often the helper's own time or patience rather than what the older adult needed \cite{5_tang2022never, 25_jin2022used, 26_korpela2024investigating}. Much of what technology caregiving demanded went on out of sight. One study named it as four kinds of work: relationship work, preparation work, facilitation work, and troubleshooting and recovery work \cite{4_kelly2024more}. A daughter checked the validity of the articles her father reposted to their family WeChat group every day: \textit{``Sometimes, I even need to read research papers so that my father can trust me''} \cite{5_tang2022never}. At its heaviest the burden was emotional and interwoven with the relationship itself. An adult child trying to stop their mother from being defrauded again described years of conflict over the issue: \textit{``I've been tormented for years. I felt utterly helpless and disappointed. Most of our fights have revolved around these issues''} \cite{36_deng2025auntie}.

\textbf{Formal helpers.} As with informal helpers, technology caregiving piled on top of what formal helpers were already doing, but what it came on top of was a job, not a life. Some care workers in assisted living facilities described technology caregiving as an additional role folded into their already multifaceted duties and demanding schedules, a burden worsened by staffing shortages \cite{21_wu2026providing}. Furthermore, help requests often arrived concurrently rather than one at a time: \textit{``sometimes you can have like five people calling you at the same time. It can be really draining and challenging''} (medication assistant in assisted living, \cite{21_wu2026providing}); \textit{``The pressure of serving all customers in line is high, so it is challenging to be always patient and meticulous''} (bank employee, \cite{25_jin2022used}). What distinguished their burden from that of informal helpers was a boundary to argue with. Formal helpers could and did contest whether the technology caregiving work was theirs at all: \textit{``A lot of people don't want to stop and facilitate that [...] Or they don't feel like it's their job to do''} (staff member in a long-term care home, \cite{29_garnett2026perceived}). Composure was a requirement of the role---\textit{``you have to always be calm, even when you are not pleased with some behaviors of the residents''} (care assistant in assisted living, \cite{21_wu2026providing})---and among frontline staff in Chinese public services, the slightest impatience could draw a complaint \cite{22_wu2026doesn}. Institutions could also absorb what individuals could not: a social security office borrowed four to six staff from other departments during its September peak, and a hospital hired temporary nurses to guide older patients \cite{22_wu2026doesn}. Informal helpers had none of these options. They had no role to invoke when they withheld support, and the reasons they gave were personal, such as a problem that was beyond their own knowledge \cite{20_zhang2025wepilot}. Older adults were sometimes the helpers themselves. One described the difficulty of diagnosing a friend's problem at a distance: \textit{``Occasionally, my friend calls me for assistance with payment errors, but it is challenging to both understand the problem and provide a solution''} (older adult helping a friend, \cite{1_seo2023power}).

\textbf{Older adults. }Seeking help often came with a sense of imposing on people older adults depended on: \textit{``Of course, I wouldn't want to be someone who always bothers other people''} (older adult, \cite{26_korpela2024investigating}). For others it was shame at having to ask, \textit{``I'm afraid people will say that at such an old age, I can't even do this, which makes me feel embarrassed''} (older adult, \cite{17_fang2026guideme}), or fear of looking ``stupid'' or ``uninformed'' \cite{28_haque2025envisioning}. Across articles, older adults sometimes attributed such feelings to how helpers reacted \cite{31_ma2023motivation}: \textit{``If I repeat to ask questions to my son, he will definitely get impatient and sometimes shout at me `Why can't you do such an easy work?' and refuse to teach me again''} (older adult, \cite{27_hu2026not}). Such feelings shaped whether help was sought at all. Some stopped asking: \textit{``When I used to ask him for help, I often had to wait until he finished work ... I try not to bother him anymore''} (older adult, \cite{20_zhang2025wepilot}). Some rationed their requests, saving up several problems before approaching a helper again so as not to overburden a friend \cite{27_hu2026not}, or asking family members less often because they were busy too \cite{9_jin2024exploring}. Some gave up the task: \textit{``Sometimes I tried it by myself because I don't want to bother others. However, even learning from others, they can't understand my frustration. So I may even choose to abandon the function''} (older adult, \cite{27_hu2026not}).

\textbf{Determinants. }Several things influenced burden on both sides. One study reported that frequency of help-seeking increased caregiver burden \cite{20_zhang2025wepilot}. Older adults readily asked their grandchildren for help but held back with their own children \cite{8_luijkx2015grandma}; friends and peers were easier to ask \cite{27_hu2026not}. When a problem was not urgent, some wrote out a list of questions and asked them all at the same time \cite{5_tang2022never}. Setting things up in advance both increased and reduced burden. Written steps \cite{5_tang2022never, 25_jin2022used}, a helper-made interactive tutorial \cite{2_jin2022synapse}, a configured voice assistant \cite{32_zubatiy2021empowering}, and a reusable record of past requests \cite{20_zhang2025wepilot} all traded repeated support work for work done once, sparing the helper a second engagement and the older adult a second request \cite{20_zhang2025wepilot, 6_tanprasert2024helpcall}. But that work, in turn, increased the initial burden of the helper \cite{32_zubatiy2021empowering, 13_barber2025beyond}. Cultural norms sometimes obliged the helper: \textit{``It's my responsibility to help my spouse learn to use the smartphone thing because we are a family''} (older adult helping a spouse, \cite{31_ma2023motivation}). Elsewhere they dictated whether asking was acceptable at all. Few older adults were reluctant to ask younger relatives they judged ``xiaoshun,'' though they still tried to relieve pressure on those who were working or studying \cite{5_tang2022never}. Not insisting on digital independence was also reported as culturally acceptable for both parents and children \cite{22_wu2026doesn}. But another study in the same setting concluded that Chinese traditional values make older adults cautious about seeking help \cite{27_hu2026not}. The same norms could also add to the burden of the helper: \textit{``[T]hey hit you with accusations of being unfilial or remind you of all the money they spent raising you to manipulate you with guilt. It's truly infuriating''} (adult child, \cite{36_deng2025auntie}).

\textbf{Not a burden. }Technology caregiving was sometimes reported as rewarding. Care workers drew meaning and fulfillment from residents' gratitude \cite{21_wu2026providing}, and a family caregiver framed helping as repayment for a lifetime of being cared for \cite{19_piper2016technological}. In both, the reward came from the relationship rather than the task, as did much of the burden. For an older adult teaching his spouse, it came from the role itself, \textit{``a sense of accomplishment ... as if I had become a teacher''} \cite{31_ma2023motivation}. But the relationship did not only make the support possible; support was sometimes the reason for contact. Older adults valued the support episode itself, and a tool that removed the burden removed the episode with it: \textit{``Although the system is helpful, I miss the detailed explanations my daughter used to give''} (older adult, \cite{20_zhang2025wepilot}). In one study, older adults occasionally experienced a technology problem as a benefit, because it brought a visit: \textit{``At least they [children] come to visit when there's a reason''} (older adult, \cite{26_korpela2024investigating}).

\subsubsection{The Role of Digital Environment in Technology Caregiving.}  

Characteristics of the digital environment---internet connectivity, devices, application platforms and interfaces, system updates, and account and credential systems---were frequently reported in the context of technology caregiving (33 articles). These characteristics hindered support more often than they facilitated it: hindrance predominated in 24 articles and facilitation in five, and 31 articles reported the same characteristic doing both. Remote support became challenging when the helper had no view of the older adult's screen and was unfamiliar with the interface being described: \textit{``When instructing the steps remotely, it is not intuitive when describing the operations. Additionally, I can not see his smartphone interface that I am not familiar with, which prevents me from giving more precise feedback''} (a helper, \cite{27_hu2026not}). Support was also hindered when account credentials were lost: staff turnover took the Skype log-in details with it, and residents who had been using Skype were no longer able to \cite{23_zamir2018video}.

\textbf{Purpose of support.} The digital environment determined what the support was for (32 articles), e.g., privacy concerns, interruption management, permissions management, password management, onboarding new phones, or operating an application \cite{3_mendel2019my}. Those needs were often remedial---a breakdown, a lockout, or a charge the older adult had not agreed to: \textit{``They display prices like 9.9 yuan, but the payment page shows 99 yuan. Many don't even realize how much they've spent, and if they do find out, they don't know how to seek redress''} (a younger family member, \cite{36_deng2025auntie}). Interface vocabulary and conventions created a need for instructional support: \textit{``There are concepts that I do not always understand in a computer language that I do not speak every day. I ask for help when I do not know''} (older adult, \cite{15_mendel2021exploratory}). Delegation could also ease the helper's own burden, as when one carer described the credentials-based system as \textit{``open to abuse [. . .] but also facilitates us carers finding face-saving, time-efficient solutions for the older people we love and help to keep going''} \cite{13_barber2025beyond}. Motivational needs ran in both directions: anticipated complexity led one older adult to turn down the banking app her daughter recommended and to keep visiting the branch twice a month \cite{15_mendel2021exploratory}, while an appealing function---video calls with children and grandchildren who lived far away---changed the minds of older adults who had wanted to avoid computers in their homes \cite{8_luijkx2015grandma}. The environment could also remove a need rather than create one: helpers ended recurring password problems by setting up a fingerprint scanner, installing a password manager, or removing the passcode altogether \cite{3_mendel2019my}.

\textbf{Source of support.} The digital environment also decided who could help (28 articles). Account and credential systems were sometimes shared with informal helpers, e.g., shopping-app account access with adult children \cite{1_seo2023power}. Where a platform offered no proxy role at all, caregivers logged in as the older adult and posted on their behalf \cite{19_piper2016technological}. The same systems could rule a helper out, when asking for help meant handing over a login that exposed everything behind it: \textit{``[If Ming] went gambling [and] doesn't want anybody to know [he lost], he can't tell his daughter, `[Can] you log in [and] send this e-transfer for me.' Because she's going to see [and] it's embarrassing''} (family member, \cite{11_dai2025envisioning}). An app's password policy could route the work to a formal helper outright, as when one older adult forgot a password, had to work with a bank employee to get her money back, and never used the app again \cite{25_jin2022used}. The environment also determined whether a willing helper was competent to act: \textit{``I wasn't familiar with [that platform]. So, I had to look it up online myself before explaining it to her''} (younger family member, \cite{20_zhang2025wepilot}); and where digital services were available only in English, friends and family were relied on to translate, so the activity waited until a daughter could do so \cite{15_mendel2021exploratory}.

\textbf{Timing of support.} Support was needed across the stages of technology use (17 articles), and not always in a forward direction: system changes often moved an older adult back to an earlier stage. Setup was where a helper first entered, because the platform's own requirements could not be met alone: \textit{``My child helped me set up [the mobile shopping app and account] at first''} (older adult, \cite{1_seo2023power}). Support did not then taper off as the technology became routine, because the environment kept changing underneath it: regular updates brought new features, functions and visuals, so that existing knowledge became obsolete and the same thing had to be learned repeatedly \cite{26_korpela2024investigating, 22_wu2026doesn}. As one older adult put it: \textit{``Nowadays, the various software updates on the phone are so fast that it is difficult for me to learn''} \cite{31_ma2023motivation}. Because initial guidance seldom produced lasting skill, helpers found themselves providing support after each change: \textit{``You just try to open the apps, see the updates, and then just explain to them what the problem is and show them how they'll go about it. Then the next time they are in the same situation''} (assisted-living staff, \cite{21_wu2026providing}). An update could also invalidate what an older adult already knew, as a husband supporting his wife by video call from abroad described: \textit{``[Aoife]'s struggled with Windows 10 because they've gone very much to a visual side and a lot of the text prompts have sort of disappeared''} \cite{24_soubutts2025hidden}. The environment could equally create a new occasion for support rather than repeat an old one, as when QR codes became a requirement for nucleic acid testing and a husband taught his wife to use a smartphone, after which she began reading the notices sent to her village's WeChat group \cite{31_ma2023motivation}.

\textbf{Delivery of support.} Whether support arrived in person or at a distance, and whether it was structured or unstructured, followed from what the platform allowed (32 articles). Sometimes remote instruction gave the helper no real-time view, so a correct instruction could be defeated by similar-looking icons: \textit{``I told her to find the `three-dot menu', but she kept clicking on the settings, confused by similar-looking icons with three dots''} (younger family member, \cite{20_zhang2025wepilot}). Elsewhere, a brother installed a shared screen viewer so he could give remote support: \textit{``He made this team viewer, so . . . I can just call him or send him a mail, and he comes . . . to my computer''} (older adult, \cite{7_hunsaker2019he}). However, other platform properties sometimes pushed support back into the room: a video-calling platform's invitation-and-accept step meant that care-home staff had to be involved before a call could happen at all, as an adult child explained---\textit{``Staff involvement is crucial. I mean, you need to send them the [Zoom] invitation. They need to accept it. [My parents] have difficulty doing that''} \cite{4_kelly2024more}. The modality also decided when support arrived and in what form: an always-on messaging channel delivered requests into the middle of a working day, but when helpers could not respond at once, older adults were left waiting \cite{20_zhang2025wepilot}. When instructions were issued asynchronously, without knowledge of the digital environment in front of the older adult, they could also fail outright---\textit{``[online help] say, `do these steps and then get back to me', and it never works''} (older adult, \cite{6_tanprasert2024helpcall}).

\section{Discussion}


\subsection{Findings about the Technology Caregiving Framework}

In this paper, we argued that helping older adults with everyday digital activities is a practice better understood as care work than as IT support (Table~\ref{tab:caregiving-comparison}). We named this practice \textit{technology caregiving} (Section 3) and introduced a framework to characterize it (Section 4). We applied the framework to conduct a qualitative systematic review of empirical studies on supporting older adults in DADLs (Table~\ref{tab:study-description}).The four dimensions of the framework---purpose, source, timing, and delivery (Figure~\ref{fig:framework})---accounted for all the configurations of support reported across the corpus: no article described support that the dimensions could not place, nor did any category prove empty or redundant (Table~\ref{tab:support}). We also found that support recurred in a small number of configurations rather than taking a different form in each article (Section 5.2.2). We identified eleven kinds of mediating artifacts in the corpus, including norms governing whether help could be asked for or offered, hardware at the older adult's end additional to the device carrying the activity, and proximity itself. We expect further kinds to be identified with more research in the area. The final codebook is reported in Appendix~\ref{app:book}.

There was no evidence in the corpus that a category on one dimension implied a category on another: a purpose was not associated with a stage of use, nor with a source, beyond the common configurations reported in Section 5.2.2. Nor did we find evidence of how support for an older adult changed over time. Most studies were cross-sectional, and the few that followed participants over a period did not report support in a way that would show transitions between categories \cite{14_wan2019appmod, 16_hu2026looking, 18_hu2025surfacing, 23_zamir2018video, 24_soubutts2025hidden, 32_zubatiy2021empowering, 34_doyle2026factors}, so the framework's account of progressive and regressive transitions between stages remains untested. We similarly found no pattern in how a single episode shifted as it unfolded, although we anticipated such shifts (Section 4.3). Purpose could also differ by whose account an article reported: the same episode could be an instructional need for an older adult, but construed or met as a need for delegation (Section 5.2.3). Throughout our analysis, we centered on the older adult's perspective, noting where a finding rested on a helper's account.  Distinguishing a DADL from \textit{support} for a DADL proved most difficult in self-reliant support, where the same action could be either---using online search to judge whether a news article is credible is a DADL (Section 2.1), whereas using online search to find how to change a privacy setting is support for one, the tool and the action being identical and only the objective distinguishing them.


\subsection{Technology Caregiving vs. Traditional Caregiving}
The systematic literature review identified three ways in which technology caregiving resembled traditional caregiving. It was relational, carried most often by the relationship between the older adult and the helper (Section 5.2.2); it was ongoing rather than episodic, since the environment kept changing and the same needs returned (Section 5.2.4); and it carried the burden long documented in the caregiving literature \cite{pearlin1990caregiving, braithwaite1992caregiving} (Section 5.2.3). Technology caregivers described an objective burden of time, interruption, and repetition, together with a subjective burden of strain, helplessness, and depleted patience. As in traditional caregiving, that burden was largely invisible, fell unevenly, and coexisted with reward, since some helpers drew meaning from the relationship the work sustained rather than from the task it completed. Although technology caregiving is oriented to the older adult's independence rather than to a device, what ended an episode was frequently the helper's own time or patience rather than what the older adult still needed (Section 5.2.3).

Technology caregiving departed from traditional caregiving in two ways: the environment in which the work took place, and the recognition and support available to those who provided it. Traditional caregiving is shaped by the care recipient's condition and by the caregiver's capacity, both of which tend to change more slowly than a digital environment does. The same was not true of technology caregiving (Table~\ref{tab:caregiving-comparison}). Across the corpus, the properties that decided what help was needed, who could give it, when, and how were properties of platforms, devices, networks, and account systems---set by parties outside the relationship (Section 5.2.4). A platform could create a need where none had been, disqualify a willing helper by making help conditional on handing over a login, or invalidate what an older adult had already learned by redesigning an interface; helpers, for their part, sometimes closed a need by changing how an account was reached. This was why the work was rarely finished, and why it was needed again and again. In traditional caregiving, what changes the need is usually the person; in technology caregiving, a need could return because the technology changed, on a schedule neither party controlled.

Traditional caregiving is a recognized category: it has a name in policy, a paid workforce alongside the unpaid one, and formal supports that have accumulated around both \cite{Riffin_2017_JAGS_UnpaidCaregivers, Spillman_2014_MQ_NSOC, Muramatsu_2019_MedicaidHomeCareAides, stone2019future}. Technology caregiving had little of this in our corpus. Where formal help appeared, it usually came from people employed to do something else---care staff, bank employees, public service workers---who absorbed the work into a job defined by other duties, and only four articles described services that existed to provide it: community technology training programs \cite{5_tang2022never}, a help desk \cite{7_hunsaker2019he}, mobile phone service centers \cite{17_fang2026guideme}, and a device-distribution program with training and an ongoing helpline \cite{30_finkelstein2023older} (Section 5.2.2). Only the help desk was typical IT support (Section 3.2), organized around discrete incidents; the rest were oriented to the older adult's continued use, and so were formal technology caregiving.

Formal helpers could at least contest whether the work was theirs, and their institutions could absorb what individuals could not; informal helpers had no role to invoke and no institution behind them (Section 5.2.3). One dementia care professional, brainstorming in an expert interview, proposed that such a helper might receive a tax credit or some other reciprocal benefit, observing that \textit{``Some people just do it from the kindness of their heart [while] paying the price for it''} \cite{11_dai2025envisioning}. The same professional named what payment would not buy: \textit{``There might be somebody [who's] paid to do that, but they're not going to have the same rapport''} \cite{11_dai2025envisioning}. Seen as care work, this help raises a question that IT support never had to answer. How can a technology caregiver be compensated, trained, or relieved without displacing the relationship that made the help work?

\subsection{Limitations}
This study has several limitations. First, we did not appraise the methodological quality of the included articles, so support reported from a single interview study carries the same weight in our synthesis as support reported from a large survey. Second, we coded what the articles reported; the counts therefore describe how this literature reports technology caregiving, not how often each form of it arises in practice. Third, our synthesis centered on the framework's dimensions and did not extend to other aspects of technology caregiving, e.g., the timeliness and effectiveness of the support received.

\subsection{Implications for Policy and Practice}

With computing applications changing more rapidly than ever and the older adult population growing worldwide, we expect technology caregiving to become more prevalent with time, and the need for it arising even without the health decline that triggers traditional caregiving. Our findings provide empirical evidence for policymakers to make this invisible work visible and to explore ways of reducing the burden on both technology caregivers and care recipients. Technology caregiving is distinct from traditional caregiving, and treating the two as one can hinder capacity building and promote patchwork solutions. The corpus showed what that patchwork looked like: outside a handful of dedicated services, the work was absorbed by staff employed for other duties, and by family members who had no role to invoke and no institution behind them (Section 6.2). For technology caregiving, the equivalent infrastructure---a policy category, a paid workforce beside the unpaid one, and the formal supports built around both---was largely missing. Recognizing it as care work in its own right is what would make training pathways, compensation, and relief available to those who provide it.

The digitization of healthcare makes this pressing. As more of healthcare is delivered through portals, applications, and remote monitoring, technology caregiving will become integral to healthcare, and without policies and infrastructure in place, it will marginalize many. Our findings show why support is not settled at the point of adoption: because the digital environment keeps changing and the same needs often return. For example, a recent meta-analysis of patient portal use among older adults found that although in recent years, more older adults are trying out patient portals, that increase has not translated in continued use yet \cite{sakhnini2026patient, sakhnini2026more}. In sum, digital equity and inclusion work needs to consider not only the acquisition of devices, network connectivity, and initial digital skill training, but also technology caregiving, which is essential for continued technology use in vulnerable populations such as older adults.

\subsection{An Agenda for Future Research}

\subsubsection{Who else Needs Technology Caregiving} Although we focus on older adults, age and age-related physical or cognitive changes may not be the only things that give rise to technology caregiving needs. Other populations---adults with intellectual disabilities, individuals with low literacy, and individuals who do not read or speak the language in which a technology is available to them---may also need it. Future research can explore these populations, and new evidence will need to update the definition of technology caregiving, which we scope to older adults.

\subsubsection{How Society and Culture shape Technology Caregiving} No article in our corpus reported data collected in Africa, South America, or South Asia, and our search covered English-language publications only (Section 5). The corpus was not uniformly Western, Educated, Industrialized, Rich, and Democratic (WEIRD) \cite{linxen2021weird}---fourteen articles reported data from East and Southeast Asia, twelve of them from China---but it drew on few regions, and cultural norms governing whether help could be asked for or offered were reported only in the articles conducted in China (Section 5.2.2). We did retrieve studies about support for older adults' digital technology use in Central Asia \cite{moldagaliyeva2026exploring}, India \cite{chatterjee2026aging}, Iran \cite{bahadori2024information}, and Bangladesh \cite{ahmed19every}, but the activities supported were not DADLs, and so the studies fell outside our review (Section 2.1). How technology caregiving needs and strategies vary across cultures and socioeconomic conditions is a question this corpus can not answer, and answering it would better situate policy and intervention design.

\subsubsection{What Happens within a Technology Caregiving Episode}
Support rarely reached an older adult through one person or one artifact. Most articles described several artifacts carrying a single arrangement (Section 5.2.2), the same helper often moved between purposes within one episode (Section 5.2.2), and a second party could be needed before the first could act at all, as when care-home staff had to accept a video-call invitation before an adult child could reach their parent \cite{4_kelly2024more}. What we could not see was how these parts came together while an episode ran. The corpus rested largely on accounts given before or after the fact, through interviews, focus groups, and surveys, or on tasks set within a study session. A few articles used contemporaneous methods, like diary studies and researcher observation \cite{16_hu2026looking, 18_hu2025surfacing, 24_soubutts2025hidden}, but none was directed at this question. Understanding how technology caregiving unfolds---who is drawn in, in what order, which artifact is reached for, what is abandoned, and under what constraints of time, availability, and attention merits further research.

\subsubsection{How System Augmentations can Reduce Technology Caregiver Burden} People and computational tools were the two most common mediating artifacts in our corpus (Section 5.2.2). The first appeared as the social relationship itself and as the kinds of expertise a helper brought to it; the second was reported in thirteen articles, three of which described older adults putting ordinary software such as a search engine or a calendar for support, and ten of which described a system the researchers designed and studied. Nine of those ten augmented what the older adult already had rather than replacing any of it: the application carrying the DADL was left unchanged, and the support was delivered around it---as an overlay drawn on the live screen, as an alert about an application's behavior, or, in two studies, as guidance projected beside the phone through a head-mounted display \cite{9_jin2024exploring, 10_li2025explorar}. Most of these systems were built to spare the helper repeated work, to spare the older adult the reluctance to ask again, and to make support available at the moment of difficulty rather than when a helper was free. 

Each was evaluated against a generic baseline: a video tutorial, a search engine, a regular video call, or in-person help from a family member. None was evaluated against another system of the same kind, although they often addressed the same problems, such as locating an element in a feature-rich interface \cite{2_jin2022synapse, 6_tanprasert2024helpcall, 9_jin2024exploring, 10_li2025explorar, 17_fang2026guideme, 20_zhang2025wepilot, 33_gao2024easyask}, learning and remembering the order of steps \cite{2_jin2022synapse, 6_tanprasert2024helpcall, 10_li2025explorar, 20_zhang2025wepilot, 33_gao2024easyask}, or recovering from a wrong step where the application offered no way back \cite{2_jin2022synapse, 10_li2025explorar, 17_fang2026guideme, 20_zhang2025wepilot, 33_gao2024easyask}. Which design strategies fare better therefore remains unknown, and a review of these systems, together with comparative evaluations, would inform future design decisions. They also served mostly instructional and remedial needs during continued use, and were tested in single sessions; longitudinal research is needed to establish whether such augmentations build durable skill, whether caregiver burden falls over time, and how a helper's involvement might be streamlined rather than removed. Whether artificial intelligence (AI) advances can carry that streamlining is untested, and an assumption it would rest on---that the older adult can bring the system to the screen where the problem lies---may not hold \cite{sharifi2026helping, 33_gao2024easyask}.

\section{Conclusion}


Older adults are increasingly expected to conduct essential daily activities through digital platforms, and many rely on others to do so. We argued that this help is care work rather than IT support, named it technology caregiving, and introduced a framework that characterizes it by the purpose, source, timing, and delivery of the support, and by the artifacts through which it reaches the older adult. In a systematic review of 36 empirical articles, the framework accounted for all the support reported, which was relational, ongoing, and burdensome, as traditional caregiving is, but arose in a digital environment controlled by neither party. HCI researchers and the practitioners who design these platforms should thus consider designing both self-reliant and helper-in-the-loop support into the product, rather than leaving support as an afterthought once the product ships.

\bibliographystyle{ACM-Reference-Format}
\bibliography{references,corpus}

\clearpage
\appendix
\section*{Appendix}

\section{Keywords used for literature search}

\begin{table}[h]
\begin{tabular}{ll}
\textbf{Domain}        & \textbf{Search Terms}                                             \\ \hline
\#1 Older Adults       & ("older adult*" OR elder* OR aging OR senior*)                    \\
\#2 Digital Support    & (caregiv* OR helper* OR assistan* OR "AI Agent" OR "Voice Agent") \\
\#3 Digital Technology & (digital OR tech OR technolog*)                                   \\
\#4                    & \#1 AND \#2 AND \#3                                               \\ \hline
\end{tabular}
\caption{Keywords used for literature search}
\end{table}

\section{Additional details about the technology caregiving framework}
\label{app:framework}

Our \textbf{purpose} categories adapt an empirically derived taxonomy of IT support needs among older adults \cite{geerts2024exploring}, from which we depart in two ways: in its split between instructional and technical support, and in the breadth of delegation. The original taxonomy defined technical support as exclusively system-repair work---``not aimed at improving digital skills but at ensuring continued use''---and instructional support as assistance for learning to use technology. We depart from this distinction for two reasons. First, empirical studies show that these two needs are difficult to disentangle within a single activity \cite{sharifi2023b_seniorlearning, jiang2026help, pang2021technology}: while receiving instructional support to pay for parking online, an older adult may also need help installing an application and adjusting a location-sharing setting; while getting help to relocate a known feature after an interface update, they may incidentally learn about a newly introduced one. Second, a disruption to a familiar activity can originate on either side---a software update rearranges a smartphone's home screen (the \textit{system} breaks), or an older adult forgets how to share a photo via WhatsApp (the \textit{skill} breaks)---yet the help needed is the same in kind: repairing the \textit{breakdown} \cite{sharifi2023b_seniorlearning, sharifi2023a_techcaregiving}. We therefore define remedial support by what the help addresses---a breakdown to a familiar activity---rather than by whether the fix is system- or skill-related, and narrow instructional support to activities that are new to the older adult, rather than \textit{any} learning of a digital skill. Turning to delegation, we define it more broadly than the taxonomy's proxy use, a term it used for non-users reaching the digital world indirectly through someone with digital skills; on our definition, complete proxy use is one end of a range rather than the category itself \cite{sharifi2023a_techcaregiving, sharifi2023b_seniorlearning}.

Our \textbf{structure} categories build on the distinction between formal, non-formal, and informal learning in the lifelong learning literature, which separates institutional, credential-bearing instruction from organized instruction outside the formal system, and both from the incidental learning embedded in everyday activities \cite{la1982formal}. We set aside the boundary between formal and non-formal learning. What separates them is whether the instruction leads to a formal \textit{credential}; a college course does, a library workshop does not. Few older adults pursue a certificate for using an online application, so this boundary has little bearing on technology caregiving. We therefore call organized support of either kind structured, and incidental support unstructured. Where credential-bearing instruction does occur, as in some continuing-education digital literacy programs, it sits at the structured end alongside a senior-center workshop. The new labels also avoid confusion with the who dimension, where formal and informal already denote the source of support.

\section{Additional results from the systematic literature review}
\label{app:corpus}

\textbf{Publication activity} in our corpus concentrated in recent years: three quarters of the 36 articles (75\%, $n = 27$) appeared between 2022 and 2026, with the final year only partially covered by our search. This trend reflects the growing academic focus on supporting everyday technology use among older adults and underscores the timeliness of this review. Most articles (72\%, $n = 26$) appeared in HCI and computing venues, with ACM CHI alone accounting for 13, alongside DIS, MobileHCI, IMWUT, PACM HCI, ASSETS, ACM Multimedia, and the International Journal of Human-Computer Interaction. Six articles (17\%) appeared in gerontology and health venues, including JMIR Aging, BMC Geriatrics, and the International Journal of Lifelong Education. The remaining four (11\%) appeared in social science venues, such as Frontiers in Psychology and Technology in Society. These 36 articles reported 43 studies (Table~\ref{tab:study-description}).

Data collection sites clustered in a small number of regions. Fourteen articles (39\%) reported data collected in East and Southeast Asia---twelve in China \cite{2_jin2022synapse, 5_tang2022never, 9_jin2024exploring, 10_li2025explorar, 17_fang2026guideme, 20_zhang2025wepilot, 22_wu2026doesn, 25_jin2022used, 27_hu2026not, 31_ma2023motivation, 33_gao2024easyask, 36_deng2025auntie}, and one each in Singapore \cite{14_wan2019appmod} and South Korea \cite{1_seo2023power}---with most of these published in 2022 or later, reflecting growing research initiatives in technologically advanced yet rapidly aging societies. Another twelve articles reported data collected in North America, eight in the United States \cite{3_mendel2019my, 12_portz2019call, 16_hu2026looking, 18_hu2025surfacing, 19_piper2016technological, 21_wu2026providing, 30_finkelstein2023older, 32_zubatiy2021empowering} and four in Canada \cite{6_tanprasert2024helpcall, 11_dai2025envisioning, 28_haque2025envisioning, 29_garnett2026perceived}. Together, these two regions account for nearly three quarters of the corpus (72\%, $n = 26$). Eight articles reported data collected in Europe, three in the United Kingdom \cite{13_barber2025beyond, 23_zamir2018video, 24_soubutts2025hidden} and one each in Finland \cite{26_korpela2024investigating}, Ireland \cite{34_doyle2026factors}, the Netherlands \cite{8_luijkx2015grandma}, and Slovenia \cite{35_petrovvcivc2024categorical}, along with one conducted across Hungary, the Netherlands, and Switzerland \cite{7_hunsaker2019he}. Single articles reported data collected in Australia \cite{4_kelly2024more} and Israel \cite{15_mendel2021exploratory}. No article in the corpus reported data collected in Africa, South America, or South Asia.

\textbf{Participants.} The included studies drew on three kinds of data. Most (23 of 43 studies) collected data from older adults alone \cite{1_seo2023power, 2_jin2022synapse, 6_tanprasert2024helpcall, 7_hunsaker2019he, 8_luijkx2015grandma, 9_jin2024exploring, 10_li2025explorar, 12_portz2019call, 15_mendel2021exploratory, 16_hu2026looking, 17_fang2026guideme, 18_hu2025surfacing, 24_soubutts2025hidden, 26_korpela2024investigating, 27_hu2026not, 28_haque2025envisioning, 30_finkelstein2023older, 33_gao2024easyask, 35_petrovvcivc2024categorical}. One of these studies recruited seven older adults from five households, several of whom were technology caregivers to another member of their household \cite{24_soubutts2025hidden}. Fourteen studies collected data from both older adults and technology caregivers \cite{5_tang2022never, 11_dai2025envisioning, 13_barber2025beyond, 14_wan2019appmod, 20_zhang2025wepilot, 22_wu2026doesn, 23_zamir2018video, 25_jin2022used, 29_garnett2026perceived, 31_ma2023motivation, 32_zubatiy2021empowering, 34_doyle2026factors}, five of them recruiting the two as dyads \cite{11_dai2025envisioning, 14_wan2019appmod, 31_ma2023motivation, 32_zubatiy2021empowering}. The remaining six studies drew only on caregivers' accounts \cite{3_mendel2019my, 4_kelly2024more, 19_piper2016technological, 21_wu2026providing, 27_hu2026not, 36_deng2025auntie}; several of these described the older adults being helped in some detail, but that description came from the caregiver rather than from the older adult.

Sample sizes ranged from five older adults in a co-design workshop \cite{13_barber2025beyond} to 701 respondents in a national survey \cite{35_petrovvcivc2024categorical}, but the studies in the corpus were predominantly qualitative with a small sample: only four studies included more than 35 older adults \cite{7_hunsaker2019he, 8_luijkx2015grandma, 34_doyle2026factors, 35_petrovvcivc2024categorical}. Reported ages spanned the early 50s to 95, though most studies set their lower bound at 60 or 65. Where gender was reported ($n = 24$), samples skewed female, in several cases substantially \cite{1_seo2023power, 2_jin2022synapse, 8_luijkx2015grandma, 10_li2025explorar, 26_korpela2024investigating}; male-majority samples were fewer and their margins narrower \cite{15_mendel2021exploratory, 17_fang2026guideme, 34_doyle2026factors}. Technology caregivers were usually younger, with reported ages beginning at 18 \cite{3_mendel2019my, 14_wan2019appmod, 27_hu2026not}, and were most often adult children and grandchildren \cite{4_kelly2024more, 5_tang2022never, 11_dai2025envisioning, 20_zhang2025wepilot, 36_deng2025auntie}. Three studies departed from this pattern by drawing on older adult peers: spouses supporting one another within couples \cite{31_ma2023motivation}, spouses acting as care partners for participants with mild cognitive impairment \cite{32_zubatiy2021empowering}, and family members in a care home study who were themselves often over 65 and struggled with the technology they were expected to support \cite{23_zamir2018video}. Formal caregivers appeared less frequently and in institutionally specific roles---bank employees \cite{25_jin2022used}, public service providers \cite{22_wu2026doesn}, home care workers \cite{21_wu2026providing}, long-term care staff \cite{29_garnett2026perceived}, and triage nurses \cite{34_doyle2026factors}.

Several further characteristics were reported too unevenly to support corpus-wide claims, but the studies reporting them suggest a narrowing of who was studied. Educational attainment appeared in 15 articles and skewed toward the highly educated, including samples in which all participants had completed secondary school and most had attended college \cite{7_hunsaker2019he, 12_portz2019call, 34_doyle2026factors}; an exception was a rural Chinese sample in which no participant had studied beyond senior high school and roughly half had not progressed past primary school \cite{31_ma2023motivation}. Several studies recruited from nursing homes, care homes, and long-term care \cite{4_kelly2024more, 9_jin2024exploring, 21_wu2026providing, 23_zamir2018video, 29_garnett2026perceived}, one excluded older adults living in skilled nursing facilities \cite{12_portz2019call}, and only six reported living arrangements as a characteristic of the sample \cite{8_luijkx2015grandma, 22_wu2026doesn, 24_soubutts2025hidden, 30_finkelstein2023older, 32_zubatiy2021empowering, 35_petrovvcivc2024categorical}.

Studies were split between recruiting older adults with cognitive impairment and screening them out. Three studies deliberately recruited older adults with mild cognitive impairment (MCI) or self-reported cognitive concerns \cite{16_hu2026looking, 18_hu2025surfacing, 32_zubatiy2021empowering}, and one administered the Mini-Mental State Examination (MMSE) without excluding on its basis, reporting that a fifth of its sample had mild cognitive problems \cite{8_luijkx2015grandma}. Others screened such participants out: one excluded anyone with a dementia diagnosis \cite{12_portz2019call}, and in another, care home staff preferred to include residents without a dementia diagnosis \cite{23_zamir2018video}. Even studies that set out to include cognitive diversity required participants to have the capacity to consent independently \cite{11_dai2025envisioning, 16_hu2026looking}. Prior technology experience was reported in 13 articles: four stated it as an eligibility criterion, requiring participants to already own or regularly use a smartphone \cite{5_tang2022never, 15_mendel2021exploratory, 20_zhang2025wepilot, 24_soubutts2025hidden}; six recorded it per participant, most often as years of smartphone use \cite{2_jin2022synapse, 17_fang2026guideme, 22_wu2026doesn, 27_hu2026not, 33_gao2024easyask}; and two noted only that all participants were existing smartphone users \cite{9_jin2024exploring, 25_jin2022used}.
Two studies assessed proficiency with a validated instrument, the Computer Proficiency Questionnaire \cite{6_tanprasert2024helpcall} and the Smartphone Proficiency Questionnaire for Chinese Older Adults \cite{17_fang2026guideme}, and one deliberately sampled both internet users and non-users \cite{35_petrovvcivc2024categorical}.

\textbf{Study design and instruments.} We categorized the corpus into two broad study types: \textit{generative} studies, which sought to understand needs and practices and reported empirical data about technology caregiving ($n = 31$), and \textit{evaluative} studies, which assessed a system, program, or intervention ($n = 12$; Table~\ref{tab:study-description}). These studies featured considerable methodological breadth, ranging from a week-long study combining observation, interviews, and a modified diary method with six older adults \cite{18_hu2025surfacing} to a national telephone survey of 701 respondents \cite{35_petrovvcivc2024categorical}. Interviews were the most commonly used method, appearing in 22 of the 36 articles \cite{1_seo2023power, 2_jin2022synapse, 4_kelly2024more, 5_tang2022never, 6_tanprasert2024helpcall, 7_hunsaker2019he, 8_luijkx2015grandma, 11_dai2025envisioning, 15_mendel2021exploratory, 16_hu2026looking, 17_fang2026guideme, 18_hu2025surfacing, 21_wu2026providing, 22_wu2026doesn, 24_soubutts2025hidden, 25_jin2022used, 26_korpela2024investigating, 28_haque2025envisioning, 29_garnett2026perceived, 31_ma2023motivation, 33_gao2024easyask, 34_doyle2026factors}, often paired with think-aloud protocols \cite{1_seo2023power, 15_mendel2021exploratory, 25_jin2022used, 27_hu2026not, 28_haque2025envisioning} or with co-design and workshop sessions \cite{9_jin2024exploring, 13_barber2025beyond, 20_zhang2025wepilot}. Comparative user studies were the dominant form of evaluation, assessing a prototype against existing practice or against alternative designs \cite{2_jin2022synapse, 6_tanprasert2024helpcall, 10_li2025explorar, 14_wan2019appmod, 17_fang2026guideme, 20_zhang2025wepilot, 28_haque2025envisioning, 33_gao2024easyask}. Surveys were used both to characterize support practices at scale \cite{3_mendel2019my, 13_barber2025beyond, 27_hu2026not, 30_finkelstein2023older, 35_petrovvcivc2024categorical} and to measure usability and workload within evaluations \cite{2_jin2022synapse, 10_li2025explorar, 17_fang2026guideme, 33_gao2024easyask}, while focus groups elicited collective perceptions from older adults and their technology caregivers \cite{11_dai2025envisioning, 12_portz2019call, 19_piper2016technological, 27_hu2026not}. A smaller set of studies observed technology caregiving as it unfolded over time, through diary studies \cite{16_hu2026looking, 18_hu2025surfacing, 24_soubutts2025hidden}, longitudinal deployment \cite{32_zubatiy2021empowering}, action research in care homes \cite{23_zamir2018video}, and a six-month randomized controlled trial \cite{34_doyle2026factors}. One article analyzed naturally occurring data rather than recruiting participants, coding 124 social media posts and their comment threads \cite{36_deng2025auntie}.

\clearpage
\section{Codebook}
\label{app:book}


\renewcommand{\arraystretch}{1}

\begin{xltabular}{\textwidth}{
    P{2cm} | 
    L         
}
\caption{Codebook: for each category of the framework we give its definition, an example passage from the corpus coded into that category, and a non-example passage considered for it and coded out.}
\label{tab:codebook}\\
\toprule
\endfirsthead
\multicolumn{2}{l}{\textit{Table \thetable\ continued from previous page}} \\
\toprule
\endhead
\midrule
\endfoot
\bottomrule
\endlastfoot
\multicolumn{2}{l}{\textbf{Purpose: Motivational}} \\
\midrule
Definition & Support that fosters awareness, confidence, and willingness to engage with an everyday digital activity at all\\
Example & \textit{``Therefore, bank employees provide active guidance and advertisement to encourage and lead older adults to use digital banking tools like mobile banking apps.''} \cite{25_jin2022used} \\
Non-example & \textit{``My peers around me often learn from me about cell phone use, so I have more experience. I am happy to share these experiences with my spouse.''} \cite{31_ma2023motivation} \\
\midrule
\multicolumn{2}{l}{\textbf{Purpose: Instructional}} \\
\midrule
Definition & Support that shows how to carry out an everyday digital activity new to the older adult\\
Example & \textit{``I'm trying to teach my husband to text\ldots{} Because it--it matters with our kids''} \cite{19_piper2016technological} \\
Non-example & \textit{``My mom always asks me to teach her how to link her bank card for payments, but I refuse and tell her to let me know what she needs and I'll purchase it for her.''} \cite{36_deng2025auntie} \\
\midrule
\multicolumn{2}{l}{\textbf{Purpose: Remedial}} \\
\midrule
Definition & Support that addresses an everyday digital activity that has broken, changed, or gone wrong\\
Example & \textit{``The final set of work activities involved resolving technical problems and helping residents to recover from troubles with the software.''} \cite{4_kelly2024more} \\
Non-example & \textit{``[W]e identified a set of functions in commonly used apps, which older adults may not know how to perform. We then asked them colloquially if they had encountered any problems referencing the set of functions''} \cite{33_gao2024easyask} \\
\midrule
\multicolumn{2}{l}{\textbf{Purpose: Delegation}} \\
\midrule
Definition & Support in which another party carries out the everyday digital activity, or part of it, on the older adult's behalf\\
Example & \textit{``I would have to step in, in time to log on for her. And that's it. Time and time again, it was the same thing [with DBP online tasks] with her''} \cite{13_barber2025beyond} \\
Non-example & \textit{``In the one-click-go mode, the interface OAs interact with remains largely the same as in the step-by-step mode. However, OAs do not need to execute most actions themselves; the system performs these actions automatically, except for privacy-related steps, which require OAs' input.''} \cite{20_zhang2025wepilot} \\
\midrule
\multicolumn{2}{l}{\textbf{Source: Formal}} \\
\midrule
Definition & A professional, paid or unpaid, without a prior personal relationship with the older adult\\
Example & \textit{``They were also the group that was the most likely to request and successfully receive training and technical assistance from the tech support companies.''} \cite{30_finkelstein2023older} \\
Non-example & \textit{``The researchers completed all of this set up work for all participating members so that participants could simply start using the calendar and it would work. Similar setup was required for the grocery list.''} \cite{32_zubatiy2021empowering} \\
\midrule
\multicolumn{2}{l}{\textbf{Source: Informal}} \\
\midrule
Definition & Someone with an existing personal relationship: family, friend, neighbor, peer\\
Example & \textit{``Grandchildren were also willing to facilitate the use of computer devices; they demonstrated the possibilities (e.g., certain games), installed applications, and helped when necessary.''} \cite{8_luijkx2015grandma} \\
Non-example & \textit{``In response, his wife respected his space by intentionally avoiding walking on the same floor while he was in the videoconference.''} \cite{18_hu2025surfacing} \\
\midrule
\multicolumn{2}{l}{\textbf{Source: Self-reliant}} \\
\midrule
Definition & What the older adult does for themselves, without another person in the moment\\
Example & \textit{``Some participants specified that they addressed technical problems through Internet searches. For example, when asked why he does not need help, one participant (72, male, Netherlands) replied, `Because I can usually solve it myself, by looking it up on the Internet.'\,''} \cite{7_hunsaker2019he} \\
Non-example & \textit{``I found it a lot easier just to group them together and then make one big Excel file that I can add to every month. So it was a lot of time sort of flicking through up front, but it's something I can come back to every month and just pop the new bills into now.''} \cite{24_soubutts2025hidden} \\
\midrule
\multicolumn{2}{l}{\textbf{Timing: Pre-adoption}} \\
\midrule
Definition & The stage at which the intention to use a technology is formed\\
Example & \textit{``Then my son-in-law said, `If your mother wants a tablet, I will join her to the store to buy one'. (Female, 77 years, living alone)''} \cite{8_luijkx2015grandma} \\
Non-example & \textit{``Arm 1 and 2 participants received a suite of connected devices, including a blood pressure monitor, weight scale, smartwatch, and a tablet preloaded with the custom-built ProACT App''} \cite{34_doyle2026factors} \\
\midrule
\multicolumn{2}{l}{\textbf{Timing: Onboarding}} \\
\midrule
Definition & The initial setup that makes a technology usable\\
Example & \textit{``My mother needed to install PayPal, but she is very technologically illiterate, so she needed help with the installation and setup of the app.''} \cite{3_mendel2019my} \\
Non-example & \textit{``A similar situation occurred for Celine, who assisted her husband Chris to transition from paper-based bills to paperless billing''} \cite{24_soubutts2025hidden} \\
\midrule
\multicolumn{2}{l}{\textbf{Timing: Early use}} \\
\midrule
Definition & The period of experimentation that follows, before the technology is incorporated into daily life\\
Example & \textit{``Another participant successfully acquired skills to set up and use their new tablet by contacting technology support services and even offered assistance to their peers: `[The name of the technology service] girls were very helpful. I taught myself based on what they said. Then I helped others, about 30 to 40 people in the building! I'd go house to house.'\,''} \cite{30_finkelstein2023older} \\
Non-example & \textit{``We began the study by sharing the link to our prototyped probe over Zoom chat and explaining the different elements of the social media interface.''} \cite{28_haque2025envisioning} \\
\midrule
\multicolumn{2}{l}{\textbf{Timing: Continued use}} \\
\midrule
Definition & The technology has been accepted and become routine\\
Example & \textit{``During this stage, younger family members continued to provide instructions if needed and often checked if older adults had difficulties with technology use.''} \cite{5_tang2022never} \\
Non-example & \textit{``In the choice of apps, we chose less popular ones and made sure participants had never used them before the experiment.''} \cite{2_jin2022synapse} \\
\midrule
\multicolumn{2}{l}{\textbf{Structure: Structured}} \\
\midrule
Definition & Support that is planned and deliberate, organized for the purpose of providing help\\
Example & \textit{``What [Terry] does is she waits for me to send her a Zoom invite. She opens the Zoom meeting on her laptop and takes it from her office down to mum and dad's room. I always schedule the meetings for the end of the day.''} \cite{4_kelly2024more} \\
Non-example & \textit{``P3 mentioned that if family members are not around, he tends to remember the problems and asks when they return home, but declining memory becomes an issue: `I often forget those questions, and waiting for someone to come back home to ask is too troublesome.'\,''} \cite{33_gao2024easyask} \\
\midrule
\multicolumn{2}{l}{\textbf{Structure: Unstructured}} \\
\midrule
Definition & Support that is casual and spontaneous, arising in ordinary interaction\\
Example & \textit{``Grandchildren were also willing to facilitate the use of computer devices; they demonstrated the possibilities (e.g., certain games), installed applications, and helped when necessary.''} \cite{8_luijkx2015grandma} \\
Non-example & \textit{``In response, his wife respected his space by intentionally avoiding walking on the same floor while he was in the videoconference.''} \cite{18_hu2025surfacing} \\
\midrule
\multicolumn{2}{l}{\textbf{Modality: Collocated}} \\
\midrule
Definition & Support given with both parties physically present\\
Example & \textit{``The most popular technique the helpers used to provide assistance was through face-to-face interaction. In 166 of the 187 of participant stories, help was provided through face-to-face, 21 of 187 reported using phone, messaging application or in other form.''} \cite{3_mendel2019my} \\
Non-example & \textit{``Married/partnered non-users had 2.996 greater odds of asking their proxy users for help compared to the reference category (widowed, separated/divorced).''} \cite{35_petrovvcivc2024categorical} \\
\midrule
\multicolumn{2}{l}{\textbf{Modality: Remote}} \\
\midrule
Definition & Support given at a distance, synchronous or asynchronous\\
Example & \textit{``My brother has the team viewer with me, because he's living [a long distance from participant]. And in the beginning I called him daily, and he had to always come [laughs] and he then gave up. He made this team viewer, so . . . I can just call him or send him a mail, and he comes . . . to my computer.''} \cite{7_hunsaker2019he} \\
Non-example & \textit{``For example, one early morning she found there was no signal for internet TV programs. She waited until after 8:30 am to call the technician''} \cite{22_wu2026doesn} \\
\midrule
\multicolumn{2}{l}{\textbf{Modality: Hybrid}} \\
\midrule
Definition & One episode spanning a remote leg and a collocated leg; the two legs need not involve the same helper\\
Example & \textit{``Sometimes now we'll actually call on our phone and get one of the nurses to go to his room and help him wake up his computer and tell him that we're about to call. And so when we call, then he can answer and then we can see him.''} \cite{4_kelly2024more} \\
Non-example & \textit{``These online updates provide an important source of encouragement for care recipients, and caregivers relay this social information and support back to care recipients by reading or showing them posts.''} \cite{19_piper2016technological} \\
\midrule
\multicolumn{2}{l}{\textbf{Artifact: Social relationships}} \\
\midrule
Definition & The relationship itself carries the help: availability, trust, burden, reluctance to ask. \\
Example & \textit{``I mainly learn from my friends because we can communicate freely, empathize, and understand each other.''} \cite{27_hu2026not} \\
Non-example & \textit{``It's hard for me ....other staff here are really busy and if they don't really know how to use this they won't bother much...it's too much to have to learn while doing other things''} \cite{23_zamir2018video} \\
\midrule
\multicolumn{2}{l}{\textbf{Artifact: IT expertise}} \\
\midrule
Definition & Expertise in the technology itself, where that expertise is a condition of the support\\
Example & \textit{``Staff training was provided on how to use Skype. (3) Implementation- staff assisted older people to use Skype with family.''} \cite{23_zamir2018video} \\
Non-example & \textit{``My grandson, he is eleven, will support me. He really knows how to use it. He knows how to search for apps.''} \cite{8_luijkx2015grandma} \\
\midrule
\multicolumn{2}{l}{\textbf{Artifact: Domain expertise}} \\
\midrule
Definition & Expertise in the domain of the everyday digital activity: shopping, banking, health, credibility\\
Example & \textit{``Institutional staff acted as critical backstops in highly regulated and sensitive domains of social security and banking.''} \cite{22_wu2026doesn} \\
Non-example & \textit{``Then somebody will have to run down and say to the girl who was doing it. 'Could you please let these people in?' So obviously they didn't understand the technology of wait rooms.''} \cite{4_kelly2024more} \\
\midrule
\multicolumn{2}{l}{\textbf{Artifact: Aging expertise}} \\
\midrule
Definition & Knowledge about older adults and their care\\
Example & \textit{``Tech support services included initial device installation, lessons, and ongoing remote services provided by the company's staff specializing in services for older adults to support participants' tech use whenever problems arise.''} \cite{30_finkelstein2023older} \\
Non-example & \textit{``[L]obby managers have more diverse and extensive experience in helping older adults, which is the reason that we choose to interview lobby managers instead of bank tellers.''} \cite{25_jin2022used} \\
\midrule
\multicolumn{2}{l}{\textbf{Artifact: Multilingual expertise}} \\
\midrule
Definition & The helper's command of a natural language the older adult does not read or speak, where that competence carries the support\\
Example & \textit{``The problem is in English, which I understand less. Afraid to click and to use the system. . ., I wait for my daughter to translate''} \cite{15_mendel2021exploratory} \\
Non-example & \textit{``[S]ome older adults said they struggled to use voice assistants because they could not speak standard Mandarin well (e.g., O6 and O16). For O6, although the basic use of voice assistants (Figure 3d) was relatively easy, she noted that the assistants she used could not understand dialects in Shanghai.''} \cite{5_tang2022never} \\
\midrule
\multicolumn{2}{l}{\textbf{Artifact: Computational tools}} \\
\midrule
Definition & A computing application or algorithm that mediates support for an everyday digital activity \\
Example & \textit{``GuideMe's in-situ highlight served as a visual anchor, filtering out distractions. U16 emphasized that the system `points out the glowing spot' allowing for instant recognition, significantly reducing the mental effort and visual search load required to map instructions to the UI.''} \cite{17_fang2026guideme} \\
Non-example & \textit{``Smartphones with accessible design (e.g., with physical buttons, large screen display, large fonts) and intelligent voice assistants were often mentioned as the most appropriate devices for older adults due to their ease of use (e.g., Y2, Y13, Y14, Y15, and Y18).''} \cite{5_tang2022never} \\
\midrule
\multicolumn{2}{l}{\textbf{Artifact: Physical tools}} \\
\midrule
Definition & A physical (non-digital) item that mediates support for an everyday digital activity \\
Example & \textit{``I followed the bank employee's instructions to walk through the process step by step. However, if I didn't write it in my notebook, I would forget about the steps.''} \cite{25_jin2022used} \\
Non-example & \textit{``The help-giver starts Synapse on his phone, which runs in the background, and starts to demonstrate how to use App A to complete the task step-by-step. He could also add voice instructions to clarify the steps. Synapse records his interactions and his voice instructions as a tutorial script.''} \cite{2_jin2022synapse} \\
\midrule
\multicolumn{2}{l}{\textbf{Artifact: Established routines}} \\
\midrule
Definition & An arrangement that allocates helpers, places, or tasks so as to make the everyday digital activity possible. Created by older adults and by caregivers alike\\
Example & \textit{``Kai's videoconferences often occurred later in the day, and involved discussing daily plans. She created bullets in her journal with key points that she could use later in the day: `In the early morning, when my mind is still relatively clear and before anyone comes to find me, I quickly write them down.' She kept this journal within arm's reach to support recall during the call.''} \cite{16_hu2026looking} \\
Non-example & \textit{``They wanted to keep in charge of it. . . . [Every week] they would walk their bills to the bank, . . . and the bank would . . . make sure that they drew the funds out to pay the bill.''} \cite{11_dai2025envisioning} \\
\midrule
\addlinespace
\multicolumn{2}{l}{\textbf{Artifact: Cultural norms}} \\
\midrule
Definition & Norms governing whether support can be sought or given\\
Example & \textit{``[F]ew older adults were reluctant to ask younger family members for help because they thought their adult children or grandchildren are `xiaoshun' (e.g., O6, O15, O16 and O20)''} \cite{5_tang2022never} \\
Non-example & \textit{``Furthermore, participants felt that digital banking represented a societal trend and they needed to keep up with the fast-changing world and not become obsolete.''} \cite{25_jin2022used} \\
\midrule
\multicolumn{2}{l}{\textbf{Artifact: Additional devices}} \\
\midrule
Definition & Hardware at the older adult's end, additional to the one carrying the activity, that carries support to them or is the route by which a helper is reached\\
Example & \textit{``The ExplorAR was developed by Unity 1 and is displayed through a Hololens 2 headset. Additionally, we integrated the Mixed Reality Tool Kit (MRTK 3 2) to perform hand tracking and build user interface, ensuring seamless user interaction within the AR interface.''} \cite{10_li2025explorar} \\
Non-example & \textit{``He made this team viewer, so . . . I can just call him or send him a mail, and he comes . . . to my computer.''} \cite{7_hunsaker2019he} \\
\midrule
\multicolumn{2}{l}{\textbf{Artifact: Physical proximity}} \\
\midrule
Definition & Help arriving because someone is physically there rather than because of who they are\\
Example & \textit{``[T]hey've often come to us as neighbours because [we] could walk over [and] look at it with them. So we've become their technical support in many cases. [They are] 90 years old, so it might be the ability to be physically there to show somebody, as opposed to over the phone''} \cite{11_dai2025envisioning} \\
Non-example & \textit{``Third, the helper stands beside the older adult, who holds the phone, and points to the corresponding UI element with their finger, explaining the action and the reason for clicking it.''} \cite{17_fang2026guideme} \\
\end{xltabular}

\end{document}